\documentclass[a4paper,11pt]{article}
\pdfoutput=1 

\usepackage{jheppub} 

\usepackage[T1]{fontenc} 

\usepackage{mathrsfs}
\usepackage{mathtools}
\usepackage{epstopdf}
\usepackage{setspace}
\usepackage{dsfont}
\usepackage{latexsym}
\usepackage{psfrag}
\usepackage{booktabs}
\usepackage{bbm}
\usepackage{color}
\usepackage{lscape}
\usepackage[caption=false]{subfig}
\usepackage{cleveref}    
\usepackage[normalem]{ulem}

\usepackage{tikz}

\def\cint#1{\int\limits_{-i\infty}^{i\infty}\frac{\dd #1}{2\pi i}}
\def\Wtadpole{W_{\text{tad}}}
\def\Itadpole#1{\mathcal{I}^{\text{tad}}_{#1}}
\def\Wscalar{W_{\text{scalar}}}
\def\Iscalar#1{\mathcal{I}^{\text{scalar}}_{#1}}
\def\edge{\mathbf{e}}

\def\Wtwobubble{W_{\text{2-bub}}}
\newcommand{\Itwobubble}[2]{\mathcal{I}^{\text{2-bub}}_{#1,#2}}

\def\Wthreebub{W_{\text{3-bub}}}
\def\Wfourbub{W_{\text{4-bub}}}
\newcommand{\Ifourbub}[3]{\mathcal{I}^{\text{4-bub}}_{#1,#2,#3}}

\Crefname{figure}{figure}{figures}

\def\dd{\text{d}}
\def\AdS{\text{AdS}}
\def\bAdS{\partial\text{AdS}}

\DeclareMathOperator*{\res}{Res}

\title{Difference Equations and Integral Families for Witten Diagrams}

\author[a]{Mark Alaverdian,}
\author[b,c]{Aidan Herderschee,}
\author[a]{Radu Roiban}
\author[a]{and Fei Teng}

\affiliation[a]{ Department of Physics, Pennsylvania State University, University Park, PA 16802, USA}
\affiliation[b]{ Institute for Advanced Study, Einstein Drive, Princeton, NJ 08540, USA}
\affiliation[c]{Leinweber Center for Theoretical Physics, Randall Laboratory of Physics\\ The University of Michigan, Ann Arbor, MI 48109-1040, USA}

\emailAdd{mxa895@psu.edu}
\emailAdd{aidanh@ias.edu}
\emailAdd{radu@phys.psu.edu}
\emailAdd{fei.teng@psu.edu}

\abstract{
We show that tree-level and one-loop Mellin space correlators in anti-de Sitter space obey certain difference equations, which are the direct analog to the differential equations for Feynman loop integrals in the flat space. Finite-difference relations, which we refer to as ``summation-by-parts relations'',
in parallel with the integration-by-parts relations for Feynman loop integrals, are derived to reduce the integrals to a basis. We illustrate the general methodology by explicitly deriving the difference equations and summation-by-parts relations for various tree-level and one-loop Witten diagrams up to the four-point bubble level.
}

\begin{document}

\maketitle




\section{Introduction}\label{s:intro}

With the discovery of AdS/CFT correspondence \cite{Maldacena:1997re, Witten:1998qj, Gubser:1998bc}, Anti-de Sitter space (AdS) has taken a special place in high energy physics. 
Boundary correlation functions in the bulk are the most natural observables~\cite{Witten:1998qj} and, through the duality, they offer insight into the strong coupling limit of correlation functions of (gauge-invariant) operators in the boundary theory. The bulk evaluation of boundary correlation functions in terms of Witten diagrams~\cite{Witten:1998qj} follows closely that of flat space correlation functions in terms of Feynman diagrams, while accounting for the non-vanishing curvature of AdS space. 

The position dependence of the AdS metric makes the direct evaluation of boundary correlators quite involved already at tree level, see e.g.~\cite{DHoker:2002nbb} for a review, requiring nontrivial integration both in global coordinates and in the Poincar\'e patch. 
Transformation to Mellin space \cite{Mack:2009mi,Mack:2009gy,Penedones:2010ue,Fitzpatrick:2011ia,Paulos:2011ie} alleviates this problem at tree level by exposing a certain similarity between Mellin space correlators and momentum-space scattering amplitudes
in flat space. However, the transformation back to position space remains a final nontrivial step. 
On the other hand, various approaches --- e.g. position space, AdS momentum, Mellin state, bootstrap from the dual boundary theory --- led to many examples of loop-level AdS boundary correlators, see e.g.~\cite{Alday:2019nin,Alday:2021ajh, Drummond:2022dxw, Heckelbacher:2022fbx,Alday:2022rly,Aprile:2022tzr,
Giombi:2017hpr,Cardona:2017tsw,Beccaria:2014qea,Aharony:2016dwx,Carmi:2024tzj,Carmi:2021dsn}. 
Despite this progress, a systematic approach to evaluate these correlators remains elusive. 

In parallel developments, significant progress has been made in evaluating Feynman integrals in flat space scattering amplitudes.
A thread behind many of the recent developments is that, for a fixed number of external particles and loop order, all integrals that can appear regardless of the structure of the Lagrangian can be written as linear combinations of some number of \emph{master integrals}. Integration-by-parts relations~\cite{Tkachov:1981wb,Chetyrkin:1981qh}, i.e. zeroes written as integrals of total derivatives of integrands of (multiples of) Feynman integrals,  and the Laporta algorithm~\cite{Laporta:2000dsw} provide an algorithmic way to construct the reduction of integrals to a basis in dimensional regularization.
Moreover, differential equations obtained by taking derivatives of the master integrals with respect to external parameters and reducing the result to the master integral basis~\cite{Kotikov:1990kg,Remiddi:1997ny,Gehrmann:1999as}, perhaps in canonical form~\cite{Henn:2013pwa}, 
lead to analytic expressions for the master integrals.
See Ref.~\cite{Weinzierl:2022eaz} for a recent review.
These ideas have been extended to curved space calculations, primarily to the calculation of cosmological correlators, see e.g. Refs.~\cite{Arkani-Hamed:2023kig,Arkani-Hamed:2023bsv,De:2023xue,Chen:2023iix,Fan:2024iek,Grimm:2024mbw}.

It is thus interesting to explore whether the techniques developed for evaluating Feynman integrals can shed light on AdS loop integrals. In this paper, we will focus on AdS Mellin amplitudes, which share some important properties with flat space scattering amplitudes.
We will discuss how integrals determining the Mellin amplitudes can be reduced to a basis and subsequently how analytic relations among the basis elements can be found. 
The discreteness of the spectrum of AdS field excitations suggests that observables are naturally written as (infinite) sums. Thus, instead of constructing relations between integrals as vanishing integrals of total derivatives, it may be more natural to identify such relations as sums of finite-difference operators acting on appropriate summands. 
We will refer to them as {\em summation-by-parts} (SBP) relations. 
A further indication that this may be a natural framework is to recall that the known analytic results for AdS Mellin space loop integrals are given in terms of Euler (di)gamma functions and (generalized) hypergeometric functions~\cite{Penedones:2010ue,Aharony:2016dwx,Giombi:2017hpr,Carmi:2018qzm, Alday:2019nin, Meltzer:2019nbs}. Such families of functions naturally obey difference equations in their parameters.
By extension, we may then expect that more general AdS integrals would obey difference equations in their parameters --- operator dimensions for boundary and internal fields, Mellin variables, etc.\footnote{Interestingly, difference equations have been discussed also for flat space Feynman integrals, see Ref.~\cite{Laporta:2000dsw}, although not in external kinematic data but propagator exponents.} 
Unlike flat space scattering amplitudes, nontrivial integrals appear already in tree-level AdS Mellin-space amplitudes and we will verify our construction on such amplitudes. 

Our discussion will focus on individual scalar AdS/Mellin-space integrals rather than boundary correlators in a particular scalar theory.\footnote{If necessary, each such integral can be interpreted as a boundary correlator in a theory with sufficiently-many fields and sufficiently general couplings.}
We will describe a general strategy to derive difference equations and SBP relations for tree-level and loop-level integrals whose integrands are ratios of Euler Gamma functions. Integrals whose integrands can be brought to this form by introduction of suitable auxiliary integration variables are also amenable to our strategy, which we illustrate in the four-point bubble example.
However, we will not always attempt to solve the difference equations we derive. Unlike differential equations, difference equations, even with boundary conditions, only fix the result up to a periodic function. We expect that this remaining freedom can be fixed by physical constraints, which we illustrate in simple cases.
For special values of external parameters, the difference equations determine uniquely the value of the integrals as rational functions or rational combinations of Gamma functions, exposing interesting identities of hypergeometric functions at unit argument. 

In \cref{sec:tadpole} we introduce the main ideas of our approach by discussing in detail the arguably simplest example of AdS loop integral --- the tadpole. 
In \cref{sec:genstrat} we formulate our general strategy for deriving SBP relations and difference equations for certain classes of Mellin-space integrals.
In \cref{sec:scalar} we apply our construction to the tree-level four-point AdS boundary correlator due to scalar exchange, with all different dimensions. 
In \cref{sec:1loop} we discuss bubble integrals with two, three and four external lines and general choice of operator dimensions. In this section we also outline possible physical requirements that fix the remaining periodic-function ambiguity of the solution to the difference equations and also compare with existing results for the two-point and four-point bubble integrals with special choices of space-time and operator dimensions.  
We also briefly touch upon singular points of the difference equations for the four-point bubble integral and their potential consequences.
In \cref{sec:discussion} we discuss our results.
Our conventions and notation are summarized in \cref{conv}; other appendices contain the lengthy expressions of coefficients of certain SBP relations (\cref{app:sbpcoef}) and an identity that follows from AdS symmetry (\cref{identity_D12_action}). 

\section{Invitation: One-Loop Tadpole}\label{sec:tadpole}

To illustrate the derivation and use of the SBP relations, we first discuss a simple example, the one-loop AdS tadpole. Building on the lessons learned here, we will then outline the general construction in \cref{sec:genstrat}.
We will assume that the propagating fields are scalars of masses $m_i^2 = \Delta_i (\Delta_i - d)$, where $\Delta_i$ are the dimensions of the dual boundary operators.\footnote{We also set the AdS radius $R_{\text{AdS}}=1$.}

\subsection{Spectral representation}

The AdS tadpole integral for a scalar field of dimension $\Delta$ running in the loop and a $(p+2)$-point scalar derivative-free interaction is given by
\begin{equation}\label{tpdef}
\mathcal{A}_{\text{tad}}(P_i)=\int_{\AdS} \dd X\,  G_{\Delta}(X,X) \,\prod_{i=1}^{p} E_{\Delta_{i}}(X,P_{i})
\,,
\end{equation}
where $G_{\Delta}(X,Y)$ is the bulk-to-bulk propagator between points $X$ and $Y$, $E_{\Delta_{i}}$ is the $i$-th bulk-to-boundary propagator for the operator of conformal dimension $\Delta_{i}$ inserted at the boundary point $P_i$. The integration is over the bulk of $\AdS_{d+1}$.
In fact, since $G_{\Delta}(X,Y)$ only depends on the difference $(X-Y)^2$, the tadpole $G_{\Delta}(X,X)$ is independent of $X$, and thus
$M_{\text{tad}}$ factorizes as
\begin{equation}\label{tpfactor}
\mathcal{A}_{\text{tad}}(P_i)=G_{\Delta}(X,X)\int_{\AdS} \dd X\,\prod_{i=1}^{p} E_{\Delta_{i}}(X,P_{i})\equiv G_{\Delta}(X,X) \bigg[ D_{\Delta_1,\dots,\Delta_p} \, \prod_{i=1}^p C_{\Delta_i}\bigg]
\,,
\end{equation}
where we have defined the $p$-point tree-level contact diagram integral $D_{\Delta_1,\dots,\Delta_p}$ is defined in \cref{contact} and the normalization factors $C_{\Delta_i}$ are given in \cref{bBprop}.

The task is therefore to compute $G_{\Delta}(X,X)$, which is only a function of the AdS dimension $d$ and the mass of the scalar field or, equivalently, the dimension $\Delta$ of the dual operator. 
While this integral can be evaluated directly, we will take a somewhat convoluted route which will be useful for more complicated integrals.

To this end, we use the spectral representation of $G_{\Delta}(X,X)$, i.e. we start from the split representation of the bulk-to-bulk propagator, 
\begin{equation}\label{splitreptad}
G_{\Delta}(X,X)
= \cint{c} \, \frac{-2c^{2}}{\delta^{2}-c^{2}} \int_{\bAdS} \dd Q\,  E_{h-c}(X,Q)E_{h+c}(X,Q)\,,
\end{equation}
with $\delta\equiv\Delta-h$, and carry out the integral over the boundary point (see \cref{conv} for more details).
The result is (see e.g. \cite{Giombi:2017hpr})
\begin{equation}\label{spectraltp}
G_{\Delta}(X,X)
=\frac{\Gamma(h)}{4\pi^{h}\Gamma(2h)} \cint{c} \frac{2c }{\delta^{2}-c^{2}} \Wtadpole(c)\,,\qquad
\Wtadpole(c) = \frac{\Gamma(h+c)\Gamma(h-c)}{c \, \Gamma(-c)\Gamma(c)}\, ,
\end{equation}
where we introduce the notation $\Wtadpole(c)$ for convenience. We refer to $\Wtadpole(c)$ and its generalizations as the \textit{potential}. The goal is to evaluate the integral in \cref{spectraltp}. 
While this integral has been evaluated before (see e.g.~\cite{Giombi:2017hpr}), we will take a somewhat ``scenic'' route which turns out to generalize to more involved cases. To this end, it is useful to spell out the details of the integration contour and the analytic continuation needed to define it. 

For $h>0$, the potential $\Wtadpole$ diverges as $c\rightarrow\infty$, so the integral is difficult to evaluate via Cauchy's residue theorem. It is no longer so for $h<0$; we will therefore evaluate it there and analytically continue the result to positive $h$. The integration contour however is defined by the split representation (\ref{splitreptad}), which in turn assumes $h>0$. 

The function $\Wtadpole$ has two sets of poles, \emph{the right-handed poles} at $c=h+n$, and \emph{the left-handed poles} at $c=-h-n$, where $n\geqslant 0$ is an integer. For $h>0$, the two sets of poles are separated by the imaginary axis, and the integration contour is along the imaginary axis. If $h<0$, certain poles in the two families may cross the imaginary axis. We then need to deform the contour so that all left-handed poles remain to the left of the contour and all the right-handed poles are to the left of the contour.

\subsection{A basis of functions}

As the poles of the potential are integer-spaced, it turns out to 
be beneficial to recast the rational factor in \cref{spectraltp} into a different form. Thus we partial fraction it to expose the simple poles,
\begin{align}
G_{\Delta}(X,X) &=-\frac{\Gamma(h)}{4\pi^{h}\Gamma(2h)} \cint{c} \left ( \frac{1}{c-\delta}+\frac{1}{c+\delta} \right )\, \Wtadpole (c) \nonumber\\
&=-\frac{\Gamma(h)}{2\pi^{h}\Gamma(2h)}\cint{c} \frac{1}{c-\delta}\Wtadpole(c)\,,
\label{TPintegralSplit}
\end{align}
where we have used the property that $\Wtadpole$ is odd in $c$.
The potential $\Wtadpole (h,c)$ exhibits unit-spaced poles and, moreover, it is mapped into itself up to a rational $c$-dependent factor under integer shifts of the integration variable,
\begin{align}
\label{Wprop}
\Wtadpole(c+1) = \frac{h+c}{c+1-h}\Wtadpole(c) \ .
\end{align}
We will use this property --- which turns out to be ubiquitous for spectral representations of AdS loop integrals and represents a shift in the integration contour --- to derive difference relations satisfied by the integral.

With the benefit of hindsight (related to the more commonly-occurring shift symmetries in later integrals), it is convenient to focus on shifts by two units, and define the finite-difference operator
\begin{equation}\label{propertiesOfD}
D^\pm_x\, [f(x)]\equiv \frac{f(x\pm 2)-f(x)}{2} \,, \quad
D^\pm_{x}[f(x)\,g(x)]=D^\pm_x[f(x)]\,g(x\pm 2)+f(x)\,D^\pm_x[g(x)] \, .
\end{equation}
Building on the observation that $\Wtadpole(h,c)\rightarrow \Wtadpole(h,c+n)$ can be realized as a shift of the integration contour, we note that such a shift will modify the denominator of $\frac{1}{c-\delta}$. It is therefore natural to consider generalizing them so that they exhibit properties similar to those of $\Wtadpole(h,c)$. We thus define 
\begin{equation}\label{propertiesOfDPi}
\Pi_{n}(x)
\equiv\frac{1}{x(x+2)\dots(x+2n-2)} 
=\frac{1}{2^{n}\left(\frac{x}{2}\right)_{n}}\,,\quad
D^+_{x}[\Pi_{n}(x)]=-n\,\Pi_{n+1}(x) \ .
\end{equation}
where $(a)_{n}$ is the Pochhammer symbol, and $\Pi_1(c-\delta) = \frac{1}{c-\delta}$.

Thus, a family of integrals, denoted as $\Itadpole{n}$, which utilize the special properties of $\Wtadpole$ and $\Pi_{n}$ will be useful,
\begin{equation}\label{tadpolebasis}
\Itadpole{n}(\delta) = \cint{c} \Pi_{n}(c - \delta)\Wtadpole(c)\,,
\end{equation}
where the original tadpole integral~\eqref{TPintegralSplit} corresponds to $\Itadpole{1}$.
These integrals exhibit many properties analogous to those of flat space Feynman integrals. In particular, it is easy to see that 
\begin{equation}\label{scalessexp}
\Itadpole{0} = \cint{c} \Wtadpole(c) = 0\,,
\end{equation}
since $\Wtadpole$ is odd in $c$.
For generic $\Itadpole{n}$ with $n\leqslant 0$, one may see from direct evaluations that they vanish identically if $h$ is sufficiently negative, and we define their values for positive $h$ by analytic continuation. It can be understood from the SBP relation derived below and is similar in spirit to the vanishing of scaleless integrals in flat space.

\subsection{A difference equation for the tadpole and summation by parts relations}

The main observation leading to a finite difference equation for the tadpole integral is that a shift in $\delta$ can be compensated in $\Pi_1(c-\delta)$ by a shift in $c$ which may then be further processed using properties of the potential. 
Thus, upon using \cref{propertiesOfDPi}, 
\begin{align}
\label{tadpoleDifferenceEq}
D^{-}_{\delta} \big[\Itadpole{1}\big] = \cint{c} D^{-}_{\delta}\big[\Pi_1(c - \delta) \big] \Wtadpole(c) 
= - \Itadpole{2} \ .
\end{align}
We will get a difference equation for $\Itadpole{1}$ if we find a relation between $\Itadpole{2}$ and $\Itadpole{1}$. That relation will come from utilizing summation-by-parts.\footnote{
It would be interesting to explore if such a strategy may be applicable to Mellin-Barnes integrals for flat-space Feynman integrals.
}

With suitable boundary conditions, one-dimensional integrals of total derivatives of real functions vanish identically. Similarly, with suitable boundary conditions, Cauchy's theorem implies that complex-plane integrals on two infinitely-long, oppositely-oriented parallel lines also vanish identically if no poles are present between the two lines. 
Using this elementary observation, we may construct relations between the integrals $\Itadpole{1}$ and a handful of other members of the 
family \eqref{tadpolebasis}. More generically, these relations follow from  
\begin{equation}\label{eq:tpsbp}
0= \cint{c} \, D^+_{c}\,\Big[P_{c} \; \Pi_{n}(c - \delta)\,\Wtadpole(c)\Big] \ ,
\end{equation}
where $P_{c}$ is an arbitrary polynomial in $c$. 
The integral is convergent for sufficiently large and negative $h$. 
The integration contour specified earlier is such that the shift in $c$ introduced by $D_{c}^{+}$ does not change the number of left-handed and right-handed poles. Indeed, for a sufficiently-large and positive $h$, the integrand has no poles for $0\leqslant c\leqslant 2$ and, as discussed, the continuation to $h<0$ is such that no poles cross the contour.
By evaluating the finite difference operator in the integrand of \cref{eq:tpsbp} and partial-fractioning the rational coefficients 
we find the desired relations between integrals in the family~\eqref{tadpolebasis}.
We expand the integrand in \cref{eq:tpsbp} using the properties of the difference operator~\eqref{propertiesOfD}, (for convenience, we replace $\Pi_1\rightarrow\Pi_n$)
\begin{align}\label{eq:expint}
D_{c}^{+}\,\Big[P_{c}\Pi_{n}(c- \delta)\,\Wtadpole(c)\Big] &= -n\, \Pi_{n+1}(c-\delta)\,\frac{P_{c+2}(c+h) (c+h+1)}{(c-h+1) (c-h+2)}\Wtadpole(c) \\
& \quad +\frac{1}{2} \left[\frac{(c+h) (c+h+1) P_{c+2}}{(c-h+1) (c-h+2)}-P_{c}\right]\Pi_{n}(c-\delta) \Wtadpole(c) 
\,,\nonumber
\end{align}
where on the first line we used \cref{propertiesOfDPi} and also \cref{Wprop} to relate $\Wtadpole(h, c+2)$ and $\Wtadpole(h, c)$. On the second line we used the latter to evaluate $D^{+}_{c}\,[P_{c}\Wtadpole(h,c)]$.
We note the appearance of poles independent of $\delta$, which would take us outside the family \eqref{tadpolebasis}. Choosing 
\begin{equation}
\label{PchoiceTP}
P_{c}=(c - h) (c - 1 - h) 
\end{equation}
resolves this issue.

Partial fractioning \cref{eq:expint} while exposing the $\Pi_n(c-\delta)$ dependence and plugging the result in \cref{eq:tpsbp} leads to 
\begin{align}\label{eq:tadpoleSBP}
    0 &= (2h-n+1) \Itadpole{n-1} + \big[4n^2-n(2\delta+6h+5)+(\delta+2)(2h+1)\big] \Itadpole{n} \nonumber\\
    &\quad - n (\delta+h-2n)(\delta+h-2n+1) \Itadpole{n+1} \ ,
\end{align}
which implies that all $\Itadpole{n\geqslant 2}$ are linear combinations of $\Itadpole{1}$ and $\Itadpole{2}$. For $n=1$, Eq. (\ref{eq:tadpoleSBP}) leads, upon use of \cref{scalessexp}, to a relation between $\Itadpole{1}$ and $\Itadpole{2}$,
\begin{equation}\label{SBPtadpole}
\Itadpole{2}(\delta) = \frac{(\delta-1) ( 2 h - 1 )}{(\delta + h - 2) ( \delta + h - 1)} \Itadpole{1}(\delta) \ ,
\end{equation}
implying that all integrals $\Itadpole{n\geqslant 1}$ can be evaluated solely in terms of $\Itadpole{1}$.
Last but not least, for $n=0$, upon using of \cref{scalessexp} one finds $\Itadpole{-1}=0$ and consequently 
\begin{equation}
\label{eq:zerotadpole}
\Itadpole{n\leqslant 0}=0 \ .
\end{equation}

Returning to the original problem, \cref{tadpoleDifferenceEq}, in the beginning of this section, we find the desired difference equation for the tadpole integral by simply plugging \cref{SBPtadpole} into the right-hand side of \cref{tadpoleDifferenceEq}:
\begin{equation}\label{diffeqtp}
D^{-}_{\delta} \big[\Itadpole{1}(\delta)\big] = - \frac{(\delta-1) (2 h-1)}{(\delta+h-2) (\delta+h-1)}\,\Itadpole{1}(\delta)\,,
\end{equation}
and consequently
\begin{align}\label{diffeqtp2}
\Itadpole{1}(\delta-2) = \frac{(\delta-h)(\delta-h-1)}{(\delta+h-2)(\delta+h-1)}\Itadpole{1}(\delta)\,.
\end{align}
This equation determines $\Itadpole{1}$ up to an arbitrary periodic function of $\delta$ which must be determined from other considerations.\footnote{This is akin to the specification of boundary conditions for differential equations for Feynman integrals.} The overall normalization may be determined by evaluating the original integral at one value of $\delta$, for example, at $\delta=0$ or $\delta=1$. They are
\begin{equation}\label{explicitvalues}
\Itadpole{1}(0)=-\frac{\, \Gamma (h)}{2\cos (\pi h) \Gamma (1-h)}\,,\qquad
\Itadpole{1}(1)=-\frac{\, \Gamma (h+1)}{2\cos (\pi h) \Gamma (2-h)}\, .
\end{equation}
These integrals can be evaluated via the residue theorem from \cref{tadpolebasis}. The convergence condition requires $h<1/2$, and we analytically continue the results to arbitrary $h$ after integration.

It is not difficult to check that a solution to \cref{diffeqtp2} is
\begin{align}
\label{I1minusValue}
    \Itadpole{1}(\delta) 
    = -\frac{ f(\delta) \, \Gamma(h-\delta)}{2 \cos(\pi h)\,\Gamma(1-\delta-h)} \,,
\end{align}
where $f(\delta)$ is a periodic function that satisfies $f(\delta)=f(\delta+2)$ and $f(0)=f(1)=1$.
Further physical considerations are necessary to determine $f(\delta)$.

To this end, it is useful to recall that 
the high-energy modes of the scalar field in the loop probe the short-distance geometry of AdS space, so they should contribute analogously to flat space. If the mass of the scalar field is also large, then these contributions should be akin to those of high momenta to a flat-space tadpole. In particular, they should not exhibit any zeroes of poles as the mass is varied.
Thus, we may demand that the AdS tadpole integral has no zeroes or poles for any non-negative values of $\delta$.
Thus, we are looking for an $f(\delta)$ with simple zeroes at $\delta = h+n$ and simple poles at $\delta = 1-h +n$ with $n$ any non-negative integer, i.e. a linear function of $1/{\sin \pi (h+\delta)}$. The normalization and simple-zero condition determine then
\begin{align}
f(\delta) = \frac{\sin\pi(h-\delta)}{\sin\pi(h+\delta)}\,.
\end{align}
The final answer is thus
\begin{align}\label{eq:tadpoleresult}
    \Itadpole{1}(\delta) &= - \frac{\Gamma(h-\delta)}{2\cos(\pi h)\Gamma(1-h-\delta)}\frac{\sin\pi(h-\delta)}{\sin\pi(h+\delta)} = - \frac{\Gamma(h+\delta)}{2\cos(\pi h)\Gamma(1-h+\delta)} \,,\nonumber\\
    \mathcal{G}^{\text{tad}}(\delta,h) &\equiv G_{\Delta}(X,X) = -\frac{\Gamma(h)}{2\pi^h\Gamma(2h)}\Itadpole{1}(\delta) = \frac{\Gamma(h)\Gamma(h+\delta)}{4\pi^h\cos(\pi h)\Gamma(2h)\Gamma(1-h+\delta)}\,,
\end{align}
and we define the function $\mathcal{G}^{\text{tad}}$ for future convenience.
For this simple problem, we can verify the result by a direct integration,
\begin{align}
\Itadpole{1}(\delta)&= \cint{c} \, \frac{1}{c - \delta} \, \Wtadpole(c) = \sum_{n=0}^{\infty}\res\nolimits_{c=-h-n}\frac{1}{c-\delta}\Wtadpole(c)\nonumber\\
&=-\frac{\sin(\pi h)}{\pi}\sum_{n=0}^{\infty}\frac{\Gamma(2h+n)}{n!(n+h+\delta)} = - \frac{\Gamma(h+\delta)}{2\cos(\pi h)\Gamma(1-h+\delta)} \ .
\label{direct_integration}
\end{align}
We note that for odd $d$, the result~\eqref{eq:tadpoleresult}  is divergent such that additional regularization is needed. Our result agrees with the known result obtained by an explicit evaluation of the tadpole integral for a real value of the mass of the internal scalar field, see Ref.~\cite{Giombi:2017hpr}.

While the machinery we presented above is unnecessary for calculating $\Itadpole{1}$, as evidenced by the one-line direct integration in \cref{direct_integration}, and more generally $\Itadpole{n}$, it serves as an illustration and a blueprint of the general strategy for deriving the difference equations of more complicated AdS loop integrals.
We outline this strategy in the next section, and then illustrate the strategy with more nontrivial integrals.

\section{General strategy}\label{sec:genstrat}

In section~\ref{sec:tadpole}, we discussed an example of a derivation of 
a difference equation for a simple AdS integral. In this section, we describe the general procedure, which generalizes this simple example to a large class of integrals of the form 
\begin{equation}\label{tpsbp}
\mathcal{I}^{G}=\int \Bigg[\prod_{a\in\mathsf{E}(G)}\frac{\dd c_{a}}{2\pi i} \frac{1}{c_a-\delta_a}\Bigg] W_{G}(c_{i},\Delta_{\edge},s_{ij}) \,,
\end{equation}
associated to a Witten diagram $G$.
Such integrals appear in tree-level and loop-level Mellin-space correlators. In such cases, each factor of $\frac{1}{c_a-\delta_a}$ arises from an internal edge $a\in\mathsf{E}(G)$ (or bulk-bulk propagator) of the diagram $G$, and $\delta_a$ is related to the conformal weight of the propagator $a$ through $\delta_a=\Delta_a-h$. The factor $W_{G}$ is some Euler integral, which we refer to as ``the (diagram's) potential.'' It is a function of the spectral parameters $c_i$, the (vector of) conformal weights $\Delta_{\edge} = \{\Delta_1,\Delta_2,\ldots,\Delta_n\}$ of the boundary states and the $n(n-3)/2$ independent Mellin space variables $s_{ij}$ at $n$ points. 

We will focus on cases where the potential $W_G$ is a ratio of products of Euler Gamma functions. However, at the price of introducing auxiliary integrals, it is always possible to write the potential as a sum of such ratios. 

To see this, let us briefly review the construction of the Mellin representation of a scalar Witten diagram for some graph $G$. The relevant identities are well-documented in, for example, Ref.~\cite{Giombi:2017hpr}. Starting from a position-space correlator, one can obtain the corresponding Mellin amplitude through the following procedures: 
\begin{enumerate}

\item Put all the bulk-bulk propagators into the split representation~\eqref{scalarbulktobulk}, and integrate out all the bulk points using
\begin{align}
    \int_{\AdS}\dd X\prod_{i\in\mathsf{E}(X)}\frac{\Gamma(\alpha_i)}{(-2\mathsf{P}_i\cdot X)^{\alpha_i}} = \frac{\pi^h}{2}\Gamma\left[\frac{\alpha_{\mathsf{E}(X)}-d}{2}\right] \int [\dd\gamma_{ij}] \prod_{i<j\in\mathsf{E}(X)}\frac{\Gamma(\gamma_{ij})}{\mathsf{P}_{ij}^{\gamma_{ij}}}\,,
\end{align}
where $\mathsf{E}(X)$ denotes all the bulk-boundary propagators connected to the bulk point $X$, and $\alpha_{\mathsf{E}(X)}=\sum_{i\in\mathsf{E}(X)}\alpha_i$. Note that here the boundary point $\mathsf{P}_i$ can be either one of the original boundary points $P_i$, or one of the $Q_i$ coming from splitting a bulk-bulk propagator. Similarly, $\alpha_i=\Delta_i$ if the bulk-boundary propagator appears in the original diagram and $\alpha_i=h\pm c_i$ if it originates in the split representation of a bulk-bulk propagator. The Mellin variables satisfy the constraints $\sum_{j\neq i\in\mathsf{E}(X)}\gamma_{ij}=\alpha_i$.
\item Recursively integrate out the boundary points $Q_i$ introduced by the split representation of bulk-bulk propagators using
\begin{align}
    \int_{\bAdS}\dd Q \prod_{i=1}^{m}\frac{\Gamma(\alpha_i)}{(-2\mathsf{P}_i\cdot Q)^{\alpha_i}} = \pi^{h} \int [\dd \rho_{ij}] \prod_{i<j}^{m}\frac{\Gamma(\rho_{ij})}{\mathsf{P}_{ij}^{\rho_{ij}}}\quad\text{where}\quad\sum_{i=1}^{m}\alpha_i=d\,.
\end{align}
We note that the requirement $\sum_{i}\alpha_i=d$ is always satisfied by the legs $\mathsf{P}_i$ contracted with $Q$ under integration.
\item From the Mellin variables $\gamma_{ij}$ and $\rho_{ij}$ introduced in the previous steps, identify and leave only those independent Mellin variables $\delta_{ij}$ for the $n$-point correlator and integrate out the rest. Note that certain changes of variables may be necessary. This will identify the $n$-point Mellin amplitude $\mathcal{M}$ as in \cref{eq:MellinAmp}.\footnote{We note that the above construction applies to the correlators that have a Mellin representation, namely, three points and higher. For two-point functions, see discussion in \cref{sec:bub2}.} 
\end{enumerate}
Given a scalar Witten diagram specified by a graph $G$, this algorithm yields its Mellin space representation in the form 
\begin{align}\label{eq:wittendiagram}
\mathcal{M}_{G}(s_{ij}) = \frac{1}{\Gamma\left[\frac{\sum_{i=1}^{n}\Delta_i-d}{2}\right]}\int\limits_{-i\infty}^{i\infty} \Bigg[\prod_{a\in\mathsf{E}(G)}\frac{\dd c_{a}}{2\pi i}\frac{c_{a}}{c_{a}^{2}-\delta_{a}^{2}}\Bigg]W_{G}(c_{i},\Delta_{\edge},s_{ij})\, ,
\end{align}
with one spectral integral, parameterized by $c_i$, for each internal leg of the diagram.
The potential $W_{G}(c_{i},\Delta_{\edge},s_{ij})$ is an Barnes integral, which depends on the spectral parameters $c_i$, boundary weights $\Delta_{\edge}$ and Mellin variables $s_{ij}$. One can try to perform the integrals by using the Barnes lemmas and put $W_G$ into the form of a product of Euler Gamma functions. However, this is not always possible for more complicated cases and some special tricks might be necessary. For example, we introduce an auxiliary bulk-bulk propagator in the four-point bubble with only $\phi^4$ type interactions so that $W_G$ becomes a product of Gamma functions, see \cref{sec:4bubble}.\footnote{Alternatively, one may follow Ref.~\cite{Penedones:2010ue} and integrate out the bulk and auxiliary boundary points using the Schwinger representation. This will put $W_G$ into the form of an Euler integral in the Schwinger parameters. By virtue of being an Euler integral, $W_G$ obeys certain difference equations~\cite{Matsubara-Heo:2023ylc} which generalize the simple tadpole shift relation \eqref{Wprop} and can be systematically used to derive SBP relations generalizing \cref{eq:tpsbp}.}

As a last step, we partial fraction the rational factor in \cref{eq:wittendiagram},
\begin{equation}
\label{eq:PF}
\frac{c_a}{c_a^{2}-\delta_a^{2}}=\frac{1}{2}\left(\frac{1}{c_a-\delta_a}+\frac{1}{c_a+\delta_a}\right) \ ;
\end{equation}
a simple change of variables $c_a\rightarrow -c_a$ together with the $W_G$ being odd under this transformation shows that the second term gives identical contributions to the first. 
Thus, up to some constant factor, the Mellin space expression \cref{eq:wittendiagram} for a scalar Witten diagram based on graph $G$ is given by \cref{tpsbp} with potential $W_G$.

We only briefly touch upon it the concluding section and leave for future work a general analysis along these lines. In the following we will focus on the simplest example of potentials, which are given given by a ratio of products of Gamma functions. 
It is worth mentioning however that, at the expense of additional auxiliary Mellin-Barnes integrals, $W_G$ can always be given an integral representation with integrand given by a ratio of products of Gamma functions.

To evaluate integrals of the type described above we derive difference equations in their arguments, i.e. $\delta_i, \Delta_{\edge}$ and $s_{ij}$. To this end it is convenient to define
\begin{equation}\label{tpsbpXX}
\mathcal{I}_{\pmb n}^{G}=\int \Bigg[\prod_{a\in\mathsf{E}(G)}\frac{\dd c_{a}}{2\pi i} \Pi_{n_a}(c_a-\delta_a)\Bigg] W_{G}(c_{i},\Delta_{\edge},s_{ij}) \ ,
\end{equation}
where $\pmb n = \{n_1,\dots,n_{|\mathsf{E}(G)|}\}$.
The canonical generalization of \cref{tadpoleDifferenceEq} to each $\delta_a$ is
\begin{align}
\label{DeltaDiffEq}
D^{-}_{\delta_a} \big[\mathcal{I}_{\pmb n}^{G}\big] = -(n_a+1)\mathcal{I}_{\pmb n + \pmb 1_a}^{G} \ .
\end{align}
where $\pmb 1_a$ is a vector with unit entry in position $a$ and all other entries vanishing.

If the potential $W_G$ is a ratio of products of Gamma functions, it is natural to expect that, under shifts of the external dimensions $\Delta_{\edge}$ by some vector $m_{\edge}$ with integer entries,
it transforms as
\begin{align}
\label{WshiftsofDelta}
W_{G}(c_{i},\Delta_{\edge}-m_{\edge},s_{ij}) = 
\frac{U_{m_{\edge}}(c_{i},\Delta_{\edge},s_{ij})}{V_{m_{\edge}}(c_{i},\Delta_{\edge}-m_{\edge},s_{ij})}
W_{G}(c_{i},\Delta_{\edge},s_{ij}) 
\end{align}
with some polynomials $U$ and $V$. The simplest (and perhaps most useful) example of vector $m_{\edge}$ has all but one entry vanishing.
This property also applies to the shift of Mellin variables $s_{ij}$,
\begin{align}\label{WshiftsofS}
    W_{G}(c_i,\Delta_{\edge},s_{ij}-m_{s}) = \frac{U_{m_{s}}(c_i,\Delta_{\edge},s_{ij})}{V_{m_{s}}(c_i,\Delta_{\edge},s_{ij}-m_{s})} W_{G}(c_i,\Delta_{\edge},s_{ij})\,,
\end{align}
and any other parameters that enter as arguments of Gamma functions.
To turn the property into a difference equation for ${\cal I}_{\pmb n}^G$, we define the auxiliary family of integrals
\begin{align}
\label{auxiliaryfamily}
{\cal J}^G_{\pmb{m},\pmb{n}}(\pmb{s}) &= \int \Bigg[\prod_{a\in\mathsf{E}(G)}\frac{\dd c_{a}}{2\pi i} \Pi_{n_a}(c_a-\delta_a)\Bigg] 
V_{\pmb{m}}(c_{i},\pmb{s}) W_{G}(c_{i},\pmb{s}) 
= \sum_{\pmb k } C_{\pmb{m}, \pmb{n}, \pmb{k}}(\pmb{s}) {\cal I}_{\pmb k}^G \ ,
\end{align}
where we only spell out the relevant arguments of ${\cal J}$. Here, we use $\pmb{s}\equiv\{\Delta_{\edge},s_{ij}\}$ to denote the kinematic data associated with boundary states, and $\pmb{m}$ a generic shift in these kinematic data.
We then partial-fraction the product $\prod_a \Pi_{n_a}(c_a-\delta_a) V_{\pmb{m}}(c_{i},\pmb{s})$ by recursively using the relation 
%
\begin{align}
    c^{k}\Pi_{n}(c-\delta) = c^{k-1} \Pi_{n-1}(c-\delta)+c^{k-1}(\delta-2n+2)\Pi_n(c-\delta)\,,
\end{align}
and then identify the resulting integrals with the family $\mathcal{I}_{\pmb k}^{G}$.
Since $V_{\pmb{m}}$ is a polynomial, the sum on the second line of \cref{auxiliaryfamily} has a finite number of terms. 
Acting on ${\cal J}$ with $D_{\Delta_{\edge}}^-$ or $D_{s_{ij}}^{-}$ (collective denoted at $D^{-}_{\pmb{s}}$) implies that\footnote{As originally defined in \cref{propertiesOfD}, $D_{\pmb{s}}^-$ comes with a two-unit shift. However, the discussion here applies to more general shifts.}
\begin{align}
\label{diffDelta}
D^{-}_{\pmb{s}}\big[\mathcal{J}^{G}_{\pmb{m},\pmb n}(\pmb{s})\big] &=\frac{1}{2}\int \Bigg[\prod_{a\in\mathsf{E}(G)}\frac{\dd c_{a}}{2\pi i} \Pi_{n_a}(c_a-\delta_a)\Bigg] 
\Big[ U_{\pmb{m}}(c_{i},\pmb{s})-V_{\pmb{m}}(c_{i},\pmb{s}) \Big] W_{G}(c_{i},\pmb{s})  \nonumber\\
& = \sum_{\pmb k } \Big\{D_{\pmb{s}}^- \big[C_{\pmb{m}, \pmb n, \pmb k}(\pmb{s})\big] \, {\cal I}_{\pmb k}^G 
 + C_{\pmb{m}, \pmb n, \pmb k}(\pmb{s}-\pmb{m}) D_{\pmb{s}}^-\big[{\cal I}_{\pmb k}^G\big]\Big\} \ .
\end{align}
The argument of the integral on the first line can be partial fractioned and explicitly written as a sum of ${\cal I}_{\pmb k}^G$ integrals, like in \cref{auxiliaryfamily}. 
Thus, \cref{diffDelta} expresses (linear combinations of) difference relations $D_{\pmb{s}}^-[{\cal I}_{\pmb k}^G]$ as linear combinations of the integrals ${\cal I}_{\pmb k}^G$. We illustrate the above construction for $D^{-}_{\pmb{s}}$, but it can be easily applied also to $D^{+}_{\pmb{s}}$.

The steps above relate integrals ${\cal I}_{\pmb n}^G$ with different indices ${\pmb n}$; further relations between such integrals are necessary to turn \cref{DeltaDiffEq,diffDelta} into constraints on a single AdS integral or on a single Witten diagram. These are the SBP relations, generalizing \cref{eq:tpsbp}, which we will discuss shortly.

The discussion above suggests that the integrals \eqref{tpsbp} provide a(n overcomplete) basis of functions for Witten diagrams. Indeed, if this were not the case, we would not have been able to choose express the action of the difference operators $\{D^{-}_{\delta_a},D^{\pm}_{\pmb s}\}$ on ${\cal I}_{\pmb n}^G$ in terms of the same integrals. It is possible that for more general classes of integrals, for which $W_G$ is not a ratio of Gamma functions, additional integrals are needed. If additional integrals appear in the (formal) evaluation of $D^-_x[{\cal I}_{\pmb n}^G]$ for some argument $x$, then the basis of functions must be extended to include them as well and \cref{DeltaDiffEq,diffDelta} must be supplemented with the action of difference operators on the additional integrals.

We now return to the SBP relations that generalize \cref{eq:tpsbp}
and are used to transform \cref{DeltaDiffEq,diffDelta} to (possibly inhomogeneous) linear difference equations involving a single integral $\mathcal{I}_{\pmb n}^{G}$.
The same reasoning that led to \cref{eq:tpsbp} also indicates its generalization: 
\begin{equation}\label{initisbp}
0=\int \prod_{i}\frac{\dd c_{i}}{2\pi i} D^+_{c_{i}}\Big[ P(c_{i},\Delta_{\edge},s_{ij}) \Pi_{n_{i}}[c_{i}- \delta_{i}] W_G(c_{i},\Delta_{\edge},s_{ij})\Big] \ ,
\end{equation}
where $P(c_{i},\Delta_{e},s_{ij})$ is a polynomial in its arguments which is chosen such that \cref{initisbp} can be written into the form 
\begin{equation}\label{intermedstep}
0 = \sum_{\pmb m} \int P_{\pmb m}'(c_{i},\Delta_{\edge},s_{ij}) \times \prod_{i}\frac{dc_{i}}{2\pi i}  \Pi_{m_{i}}[c_{i}- \delta_{i}] W(c_{i},\Delta_{\edge},s_{ij})
\end{equation}
where $P_{\pmb m}'$ are polynomials in their arguments and the sum includes a finite number of terms. 
Such polynomials exists if $W(c_{i}+2,\Delta_{\edge},s_{ij})$ is related to $W(c_{i},\Delta_{\edge},s_{ij})$ by a multiplicative rational function of its arguments; then, $P(c_i, \Delta_{\edge}, s_{ij})$ is chosen to cancel the poles of 
$D_c^-[W_G]/W_G$, in close similarity with \cref{WshiftsofDelta,WshiftsofS,auxiliaryfamily} and \cref{eq:tpsbp,eq:expint,PchoiceTP}.\footnote{It would be interesting to explore cases in which $D_c^-[W_G]/W_G$ is not a rational function.}
Upon partial fractioning the rational factors in \cref{intermedstep}
and repeatedly using the properties of $\Pi_n(c-\delta)$, we systematically obtain the SBP relation
\begin{equation}\label{eq:sumbparts}
0=\sum_{\pmb k} a_{\pmb k}^{}(c_i, \Delta_{\edge}, s_{ij})
\mathcal{I}_{\pmb k}^G \ ,
\end{equation}
where the $a_{\pmb n}$ are meromorphic coefficients in its arguments.

Closure of these relations allows us to use them to cast the difference equations in the external parameters, \cref{DeltaDiffEq,diffDelta}, into equations for Mellin amplitudes. To this end and similarly with flat-space Feynman integrals~\cite{Laporta:2000dsw}, end-point integrals which either vanish or can be easily evaluated play an important role. Examples of the former are the integrals in \cref{scalessexp} and the subsequent discussion; we will see in the following sections examples of the latter. 
The outcome of this algorithm is a set of first order (typically) inhomogeneous finite-difference equations in external parameters for a basis of independent master integrals: 
\begin{equation}\label{differeneq}
D_{x}[\vec{\mathcal{I}}{}^{G}]=\hat{A}_{G}\cdot \vec{\mathcal{I}}{}^{G} +  {\vec B}_{G}\,, \qquad x\in \{\delta_a, \Delta_{\edge}, s_{ij}\} \ .
\end{equation}
In these equations $\vec{\mathcal{I}}{}^{G}$ stands for a vector of integrals associated to the original graph $G$. The components of this vector are labeled by the multi-component index ${\pmb n}$ in \cref{tpsbpXX}. 
Various steps above may lead to some entry of this index to become negative, i.e. certain $\Pi_n(c_i-\delta_i)$ factors exhibit no poles; as we will see in examples below, such absence of poles can be interpreted as the collapse of an AdS bulk-bulk propagator~\cite{Herderschee:2021jbi}. With sufficiently-many collapsed propagators, the integrals become easy to evaluate or may even vanish.
The inhomogeneous term ${\vec B}_G$ arises from such nonvanishing integrals
that can be evaluated and hasten the closure of the summation-by-parts algorithm. This term may also be absorbed into the homogeneous one, at the price of extending the vector ${\vec {\cal I}}^G$ to also include these simple integrals. 
For the class of integrals discussed here, the entries of the matrix $\hat{A}_G$ are rational functions. One may however envision that the entries of this matrix may be more complicated functions for more involved classes of integrals. It is beyond the scope of this paper to develop computational strategies to solve \cref{differeneq}, but we comment on possible approaches in the discussion.

We note that in a generic physical theory, boundary correlators may receive contributions from Witten diagrams corresponding to graphs $G_i$ of different topologies. It is convenient to combine the integrals and their daughters corresponding to all contributing graphs and mod out by the symmetries between them.
This operation is nontrivial because, due to propagator collapse, integrals that start out as corresponding to distinct graphs, have daughters that correspond to identical graphs. 
The result is the list of {\em global master integrals} for those correlators; these integrals are labeled by a graph $G_i$ and a multi-component index.
Thus, the analog of \cref{differeneq} is obtained by formally replacing $G\mapsto {\vec G}$, and the rows and columns of the matrix $\hat{A}_{\vec G}$ are labeled by pairs $(G_i,{\pmb n}_i)$

We also note that, for complicated integrals, the difference equations and the SBP relations are intertwined. In the process of eliminating the auxiliary integrals (namely, those integrals in the SBP relations other than the ones being evaluated), difference operators of higher orders can be generated. We will see this explicitly in \cref{sec:4bubble}.

\section{Scalar exchange}\label{sec:scalar}

\begin{figure}
    \centering
    \begin{tikzpicture}[every path/.style={very thick}]
        \pgfmathsetmacro{\r}{1.5};
        \coordinate (l) at (-\r/3,0);
        \coordinate (r) at (\r/3,0);
        \draw (0,0) circle (\r);
        \filldraw (l) circle (1pt) (r) circle (1pt);
        \draw (l) -- (r) node [below=0pt,pos=0.5] {$\Delta$};
        \draw (r) -- (30:\r) node [right=0.5pt] {$\Delta_3$};
        \draw (r) -- (-30:\r) node [right=0.5pt] {$\Delta_4$};
        \draw (l) -- (150:\r) node [left=0.5pt] {$\Delta_1$};
        \draw (l) -- (-150:\r) node [left=0.5pt] {$\Delta_2$};
    \end{tikzpicture}
    \caption{The $s$-channel scalar exchange.}
    \label{fig:scalar}
\end{figure}
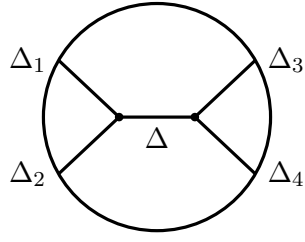
Perhaps the simplest example of an integral of the type discussed in the previous section is the single-channel contribution to the tree-level four-scalar correlation function in massive $\phi_1\phi_2\phi_3$ theory. We discuss in detail the $s$ channel as shown in \cref{fig:scalar}, 
\begin{align}
\label{eq:schannelScalar}
\mathcal{A}_{s} &= \int_{\AdS} \dd X_{1} \dd X_{2} E_{\Delta_{1}}(X_{1},P_{1})E_{\Delta_{2}}(X_{1},P_{2}) G_{\Delta}(X_{1},X_{2})E_{\Delta_{3}}(X_{2},P_{3})E_{\Delta_{4}}(X_{2},P_{4})\,,
\end{align}
where the bulk-boundary propagator $E_{\Delta_i}$ is given in \cref{bBprop}, and the split representation of the bulk-bulk propagator $G_{\Delta}$ is given in \cref{scalarbulktobulk}.
The $t$ and $u$ channels follow by relabeling of external data.

We can directly apply the procedure outlined in \cref{sec:genstrat} to compute the $s$-channel scalar exchange contribution in \cref{eq:schannelScalar}. The Mellin space amplitude was first given in Ref.~\cite{Penedones:2010ue},
\begin{align}\label{eq:Ms}
\mathcal{M}(\delta,s) = \frac{1}{\Gamma\left(\frac{\Delta_{1234}}{2}-h\right)\Gamma\left(\frac{\Delta_{12}-s}{2}\right)\Gamma\left(\frac{\Delta_{34}-s}{2}\right)}\int\limits_{-i\infty}^{i\infty}\frac{\dd c}{2\pi i}\frac{l(c,s) l(-c,s)}{(\Delta-h)^2 - c^2} 
\end{align}
where $\delta = \Delta-h$, $\Delta_{ij\ldots} = \Delta_i+\Delta_j+\ldots$ and 
\begin{align}
l(c,s) &= \frac{\Gamma\left(\frac{h+c-s}{2}\right)\Gamma\left(\frac{\Delta_{12}-h+c}{2}\right)\Gamma\left(\frac{\Delta_{34}-h+c}{2}\right)}{2\Gamma(c)} \ .
\end{align}
The Mandelstam variable $s$ is related to the Mellin variable $\delta_{12}$ through $\delta_{12} = \frac{\Delta_{12}-s}{2}$. The integration over $c$ is along the imaginary axis, and we choose the parameters such that the poles of $l(c,s)$ are all on the left hand side of the imaginary axis, from which we can obtain generic results through analytic continuation.
Upon partial fractioning of the rational part of the integrand of \eqref{eq:Ms} using \cref{eq:PF} and changing variables $c\rightarrow -c $ in the second term \cref{eq:Ms} takes, up to an overall factor,  the form \eqref{tpsbp}
\begin{align}\label{eq:Ms2}
    \mathcal{M}(\delta,s)= \frac{2}{\Gamma\left(\frac{\Delta_{1234}}{2}-h\right)\Gamma\left(\frac{\Delta_{12}-s}{2}\right)\Gamma\left(\frac{\Delta_{34}-s}{2}\right)} \cint{c} \frac{1}{c-\delta}\Wscalar(c,s)\,,
\end{align}
and identifies the potential function $\Wscalar$ for the $s$-channel scalar exchange as
\begin{align}
    & \Wscalar(c,s) = - \frac{1}{2c} l(c,s) l(-c,s)\,.
\end{align}
The asymptotic behavior of $l(c,s)l(-c,s)$ at $c\rightarrow\infty$ is
\begin{align}
    l(c,s)l(-c,s)\xrightarrow{c\rightarrow\infty}\frac{\Gamma(c/2)^3\Gamma(-c/2)^3}{\Gamma(c)\Gamma(-c)} = \frac{8\pi^2}{c^2}\frac{\sin \pi c}{(\sin\frac{\pi c}{2})^3} \,.
\end{align}
The $1/c^2$ fall-off guarantees that there are no poles at infinity and thus the contour integral in \cref{eq:Ms2} can be computed via Cauchy's theorem as a sum of residues. The result is
\begin{align}
\label{eq:direct}
    \mathcal{M}(\Delta-h,s) = \frac{{}_3F_2\big(1,1+\frac{s+\Delta-d}{2},\frac{\Delta_{1234}-d}{2};1+\frac{\Delta_{12}+\Delta-d}{2},1+\frac{\Delta_{34}+\Delta-d}{2};1\big)}{(\Delta_{12}+\Delta-d)(\Delta_{34}+\Delta-d)} \ ,
\end{align}
which agrees with eqs.~(40)-(41) of Ref.~\cite{Penedones:2010ue} for generic values of parameters.
We will show that this Mellin amplitude obeys difference equations constructed following the strategy described in \cref{sec:genstrat}.

To this end, it is useful to define\footnote{In the rest of this section, ${\cal M}$ and ${\cal I}$ have $c,s, \Delta_{12}$ and  $\Delta_{34}$ as arguments; to simplify the notation we will write explicitly only the arguments relevant for the particular equation.}
\begin{align}
    \Iscalar{n} = \cint{c} \Pi_n(c-\delta)\Wscalar(c,s, \Delta_{12}, \Delta_{34})\,,
\end{align}
such that the Mellin amplitude becomes
\begin{align}
    \mathcal{M}(\delta,s) = \frac{2 \, \Iscalar{1}}{\Gamma\left(\frac{\Delta_{1234}}{2}-h\right)\Gamma\left(\frac{\Delta_{12}-s}{2}\right)\Gamma\left(\frac{\Delta_{34}-s}{2}\right)}\,.
\end{align}
We will focus on the evaluation of $\Iscalar{1}$. The difference equations
in $\delta$, $s$, $\Delta_{12}$ and $\Delta_{34}$ are
\begingroup
\allowdisplaybreaks
\begin{align}
\label{eq:diffeq_scalarexchange}
D_{\delta}^{-}[\Iscalar{1}] &= \cint{c} D_{\delta}^{-}[\Pi_1(c-\delta)]\Wscalar(c,s) = - \Iscalar{2}\,,
\\
    D_s^{-}\big[\Iscalar{1}\big] &= \cint{c} \frac{1}{c-\delta} D_s^{-}\big[\Wscalar(c,s)\big] = \frac{(s-h)^2-\delta^2-4}{8}\Iscalar{1}-\frac{1}{8}\Iscalar{-1}\,,
\\
    \Iscalar{1}(\Delta_{12}+2) &= -\frac{1}{4}(\delta+h-\Delta_{12})(\delta-h+\Delta_{12})\Iscalar{1}(\Delta_{12}) -\frac{1}{4}\Iscalar{-1}\,,\\
    \Iscalar{1}(\Delta_{34}+2) &= -\frac{1}{4}(\delta+h-\Delta_{34})(\delta-h+\Delta_{34})\Iscalar{1}(\Delta_{34}) -\frac{1}{4}\Iscalar{-1}\,.
\end{align}
\endgroup
We thus see that, as discussed in \cref{sec:genstrat}, the key to determining $\Iscalar{1}$ is finding the SBP relations between it and 
the integral with shifted index.

The integral family governed by $\Wscalar$ has a single spectral parameter, so the SBP relation following from \cref{intermedstep} is
\begin{align}
\label{eq:SBPscalarexchange}
    0 = \cint{c} \, D_{c}^{+}\Big[P_{c}\,\Pi_{n}(c-\delta)\Wscalar(c,s)\Big] \ ,
\end{align}
where the polynomial $P_{c}$ is chosen such that $P_{c+2}$ removes the poles of the ratio 
\begin{align}
    \frac{\Wscalar(c+2,s)}{\Wscalar(c,s)} = \frac{(h-s+c)(\Delta_{12}-h+c)(\Delta_{34}-h+c)}{(h-s-c-2)(\Delta_{12}-h-c-2)(\Delta_{34}-h-c-2)} \,.
\end{align}
Thus, we choose
\begin{align}
    P_c = (h-s-c)(\Delta_{12}-h-c)(\Delta_{34}-h-c) \,,  
\end{align}
which, upon direct expansion and partial fractioning of the integrand of \cref{eq:SBPscalarexchange} leads to the scalar exchange realization of \cref{eq:sumbparts} (the summation vector $\vec k$ in that equation has a single component, so we omit the vector notation):
\begin{align}
\label{eq:SBPscalar}
    \Iscalar{n-3} + a_{n-2} \Iscalar{n-2} + a_{n-1} \Iscalar{n-1} + a_{n} \Iscalar{n} + a_{n+1} \Iscalar{n+1} = 0 \ .
\end{align}
The coefficients $a_k$ ($k=n-2,\dots,n+1$) are 
\begin{align}\label{eq:aa_scalar}
    a_{n-2} &= 3 \delta - 7 n + 12 \nonumber\\
    a_{n-1} &= \Big[\Delta_{12}\Delta_{34}-(n+s)\Delta_{1234}\Big] \nonumber\\
    &\quad +(28 - h^2 - 42 n + h n + 18 n^2 + 2 h s + n s + 18 \delta - 15 n \delta + 3 \delta^2)\nonumber\\
    a_{n} &= (\delta - 3n + 2)(\Delta_{12}\Delta_{34} - s \Delta_{1234} -h^2+2hs)+2n(2n-\delta-1)(\Delta_{1234}-h) \nonumber\\
    &\quad + (\delta-3n+2)^3 + n \Big[7 n^2 - n (18 + 4 s + 3 \delta) + 
   2 (4 + s + 3 \delta + s \delta)\Big] \nonumber\\
    a_{n+1} &= n(s+2n-h-\delta)(\Delta_{12}+\delta-2n-h)(\Delta_{34}+\delta-2n-h) \ .
\end{align}
For example, we have
\begin{align}
    \Iscalar{2} &= \frac{(\delta-1)\Big[(\Delta_{12}-s-2)(\Delta_{34}-s-2)+(\delta-h)(\delta+h-2)-s(s-2h+2)\Big]}{(\delta+h-s-2)(\Delta_{12}+\delta-h-2)(\Delta_{34}+\delta-h-2)} \Iscalar{1} \nonumber\\
    & \quad + \frac{(3\delta+5)\Iscalar{-1}+\Iscalar{-2}}{(\delta+h-s-2)(\Delta_{12}+\delta-h-2)(\Delta_{34}+\delta-h-2)} \ ,
\end{align} 
where we discarded $\Iscalar{0}$ as it vanishes identically because $\Wscalar(c,s)$ is odd in the spectral parameter $c$. The vanishing of $\Iscalar{0}$ as well as the vanishing of the integrals
\begin{align}
\cint{c} \, c^{2k}\, \Wscalar(c,s) = 0 \ ,
\end{align}
determine part of the boundary integrals. These relations also imply that 
$\Iscalar{-2}$ is related to $\Iscalar{-1}$,
\begin{align}\label{eq:Im2_scalar}
    \Iscalar{-2} &= \cint{c} \, \Pi_{-2}(c-\delta) \Wscalar(c,s) = -2(\delta+3) \Iscalar{-1}\, .
\end{align}
The vanishing of the $a_{n+1}$ coefficient in \cref{eq:SBPscalar} 
for $n=0$, decouples $\Iscalar{n<-1}$ from $\Iscalar{n>0}$ and implies that the SBP relation~\eqref{eq:SBPscalar} determines all the integrals $\Iscalar{-n}$ with negative indices in terms of $\Iscalar{-1}$. For example, if we set $n=0$ in \cref{eq:aa_scalar}, we will get
\begin{align}
    \Iscalar{-3} &= (44+h^2-2hs+24\delta+3\delta^2+s\Delta_{1234}-\Delta_{12}\Delta_{34}) \, \Iscalar{-1}
\end{align}
after using \cref{eq:Im2_scalar}. Therefore, the integral
\begin{align}\label{eq:Iscalarbdry}
    \Iscalar{-1} &= \cint{c} \Pi_{-1}(c-\delta)\Wscalar(c,s)
    =-\frac{1}{2}\int\limits_{-i\infty}^{i\infty}\frac{\dd c}{2\pi i} l(c,s)l(-c,s)\, ,
\end{align}
is the only remaining boundary integral. 

The integral~\eqref{eq:Iscalarbdry} can be directly computed through summing residues. However, here we will follow a shortcut.
Noticing that removing the denominator $(\Delta-h)^2 - c^2$ from the integrand of \eqref{eq:Ms} maps it to the Mellin amplitude for the four-point scalar contact diagram, 
\begin{align}\label{eq:4contact}
   1= \mathcal{M}_{\text{contact}} &= \frac{1}{\Gamma\left(\frac{\Delta_{1234}}{2}-h\right)\Gamma\left(\frac{\Delta_{12}-s}{2}\right)\Gamma\left(\frac{\Delta_{34}-s}{2}\right)}\int\limits_{-i\infty}^{i\infty}\frac{\dd c}{2\pi i}l(c,s) l(-c,s) \\
   &= \frac{-2\,\Iscalar{-1}}{\Gamma\left(\frac{\Delta_{1234}}{2}-h\right)\Gamma\left(\frac{\Delta_{12}-s}{2}\right)\Gamma\left(\frac{\Delta_{34}-s}{2}\right)} \,, \nonumber
\end{align}
we can immediately read off that $\Iscalar{-1}$ is given by 
\begin{align}\label{eq:Im1scalar}
    \Iscalar{-1} = - \frac{1}{2} \, \Gamma\left(\frac{\Delta_{1234}}{2}-h\right)\Gamma\left(\frac{\Delta_{12}-s}{2}\right)\Gamma\left(\frac{\Delta_{34}-s}{2}\right)\, .
\end{align}
The relation~\eqref{eq:4contact} follows from inserting a bulk delta function into the contact diagram and then converting to the spectral representation.

With these SBP relations in hand, we return to the difference equations
\cref{eq:diffeq_scalarexchange} and eliminate all the integrals different from the desired one, $\Iscalar{1}$. Slightly reorganizing all expressions
in the form of recursion relations, we find
\begingroup
\allowdisplaybreaks
\begin{align}
    \Iscalar{1}(\delta-2) &= \frac{(h-s-\delta)(\Delta_{12}-h-\delta)(\Delta_{34}-h-\delta)}{(h-s+\delta-2)(\Delta_{12}-h+\delta-2)(\Delta_{34}-h+\delta-2)} \Iscalar{1}(\delta) \nonumber\\*
    & \quad + \frac{(\delta-1)\,\Gamma\left(\frac{\Delta_{1234}}{2}-h\right)\Gamma\left(\frac{\Delta_{12}-s}{2}\right)\Gamma\left(\frac{\Delta_{34}-s}{2}\right)}{(\delta+h-s-2)(\Delta_{12}+\delta-h-2)(\Delta_{34}+\delta-h-2)}
    \\
    \label{eq:DE_s_scalar}
    \Iscalar{1}(s-2) &= \frac{1}{4}(h-s-\delta)(h-s+\delta) \Iscalar{1}(s) \nonumber\\*
    &\quad + \frac{1}{8}\,\Gamma\left(\frac{\Delta_{1234}}{2}-h\right)\Gamma\left(\frac{\Delta_{12}-s}{2}\right)\Gamma\left(\frac{\Delta_{34}-s}{2}\right)
    \\
    \Iscalar{1}(\Delta_{12}+2) &= -\frac{1}{4}(\delta+h-\Delta_{12})(\delta-h+\Delta_{12})\Iscalar{1}(\Delta_{12}) \nonumber\\*
    & \quad + \frac{1}{8}\,\Gamma\left(\frac{\Delta_{1234}}{2}-h\right)\Gamma\left(\frac{\Delta_{12}-s}{2}\right)\Gamma\left(\frac{\Delta_{34}-s}{2}\right)
    \\
    \Iscalar{1}(\Delta_{34}+2) &= -\frac{1}{4}(\delta+h-\Delta_{34})(\delta-h+\Delta_{34})\Iscalar{1}(\Delta_{34}) \nonumber\\*
    & \quad + \frac{1}{8}\,\Gamma\left(\frac{\Delta_{1234}}{2}-h\right)\Gamma\left(\frac{\Delta_{12}-s}{2}\right)\Gamma\left(\frac{\Delta_{34}-s}{2}\right)  \ ,
    \label{eq:del34shift}
\end{align}
\endgroup
where in each relation we manifested only the argument of $\Iscalar{1}$ relevant for the recursion. As expected from the general discussion in \cref{sec:genstrat}, these relations are inhomogeneous, with the inhomogeneity term given by a graph with collapsed propagators (in this case the four-point contact graph). We note that \cref{eq:DE_s_scalar} is equivalent to the Casimir equation for scalar exchange Witten diagrams discussed in Ref.~\cite{Zhou:2020ptb}.

Similarly to first-order inhomogeneous differential equations, the solution is a sum of a general solution of the homogeneous relation and a particular solution of the inhomogeneous one. The remaining freedom is then fixed by boundary conditions and possibly further physical requirements on the pole structure of $\Iscalar{1}$ or of the Mellin amplitude.

Interestingly, the first two relations take a somewhat more pleasant form when written in terms of the Mellin amplitude:
\begin{align}
\label{eq:scexc_del}
    \mathcal{M}(\delta-2,s) &= \frac{(h-s-\delta)(\Delta_{12}-h-\delta)(\Delta_{34}-h-\delta)\mathcal{M}(\delta,s)+2(\delta-1)}{(h-s+\delta-2)(\Delta_{12}-h+\delta-2)(\Delta_{34}-h+\delta-2)} \,,
\\
    \mathcal{M}(\delta,s-2) &= \frac{(h-s-\delta)(h-s+\delta)\mathcal{M}(\delta,s)+1}{(\Delta_{12}-s)(\Delta_{34}-s)} \,,
\\
    \mathcal{M}(\Delta_{12}+2) &= - \frac{(\delta+h-\Delta_{12})(\delta-h+\Delta_{12})\mathcal{M}(\Delta_{12})-1}{(\Delta_{12}-s)(\Delta_{1234}-2h)} \,,
    \label{eq:scexc_Del12}
    \\
    \mathcal{M}(\Delta_{34}+2) &= - \frac{(\delta+h-\Delta_{34})(\delta-h+\Delta_{34})\mathcal{M}(\Delta_{34})-1}{(\Delta_{34}-s)(\Delta_{1234}-2h)}  \, .
\label{eq:scexc_Del34}    
\end{align}
It is not difficult to verify that the result of the direct computation of the scalar-exchange Mellin amplitude \cref{eq:direct} obeys these recursions.

The difference equations we just derived, Eqs.~\eqref{eq:scexc_del}-\eqref{eq:scexc_Del34}, have the interesting property that, for certain values of the parameters, they determine the amplitude to be a rational function of external data. These special values, which here and below we denote with a star, are determined by the zeroes of the coefficient of ${\cal M}$ on the right-hand side of these equations. Consider for example \cref{eq:scexc_del}; for $\Delta_{34*} = \delta+h$ or $\Delta_{12*}=h+\delta$ the coefficient of $\mathcal{M}(\delta,s)$ on the right-hand side of this equation vanishes and we are left with 
\begin{align}
    \mathcal{M}(\delta-2,s)\Big|_{\Delta_{12}=\delta+h} &= 
    \frac{1}{(h-s+\delta-2)(\Delta_{34}-h+\delta-2)}  \ , 
    \\
    \mathcal{M}(\delta-2,s)\Big|_{\Delta_{34}=\delta+h} &= 
    \frac{1}{(h-s+\delta-2)(\Delta_{12}-h+\delta-2)} \ .
\end{align}
with ${\cal M}$ in \cref{eq:direct}. The same relations follow from eqs.~\eqref{eq:scexc_Del12} and \eqref{eq:scexc_Del34} for the same values of $\Delta_{12}$ and $\Delta_{34}$, respectively.
It is not difficult to verify that they are indeed satisfied. 

\section{One-Loop examples}\label{sec:1loop}

Based on the analysis of the AdS tadpole integral in \cref{sec:tadpole}, in \cref{sec:genstrat} we outlined a general difference-equation-based reduction of AdS integrals to a basis and the subsequent derivation of difference equations for the basis elements. Having illustrated this strategy with the tree-level scalar-exchange integral in the previous section,  we proceed in this section to discuss one-loop bubble diagrams with various numbers of external legs. 
Similar to the tadpole integral, in a gauge/gravity duality framework, they represent $1/N$ corrections to the anomalous dimensions of boundary operators, as well as part of the $1/N$ corrections to the three- and higher-point correlation functions. 

Unlike tadpole diagrams, increasing the number of external points of the bubble introduces additional complexity by increasing the number of kinematic variables ($\Delta_i$, $\delta_a$, Mellin variables). 
For that reason, we start with the case of the two-point bubble, for which the external kinematics is completely fixed in terms of the operator dimensions $\Delta_i$ and does not include Mellin variables. The result has been known for some time~\cite{Giombi:2017hpr} and we will verify it
here using the methods developed in~\cref{sec:genstrat}.

\subsection{Two-point bubble}\label{sec:bub2}

\begin{figure}
    \centering
    \begin{tikzpicture}[every path/.style={very thick}]
        \pgfmathsetmacro{\r}{1.5};
        \draw (0,0) circle (\r);
        \draw (0,0) circle (\r/3);
        \node at (0,\r/3) [above=0pt] {$\Delta$};
        \node at (0,-\r/3) [below=0pt] {$\Delta'$};
        \filldraw (\r/3,0) circle (1pt) (-\r/3,0) circle (1pt);
        \draw (\r/3,0) -- (\r,0) node [right=0pt] {$\Delta_2$};
        \draw (-\r/3,0) -- (-\r,0) node [left=0pt] {$\Delta_1$};
    \end{tikzpicture}
    \caption{The two-point bubble.}
    \label{fig:2bub}
\end{figure}
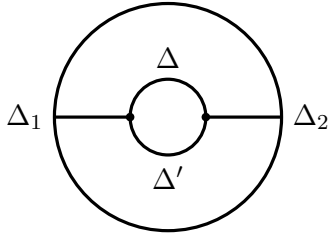

The two-point scalar bubble integral, as shown in \cref{fig:2bub}, represents the two-point correlation function in $\phi_1\phi_2\phi_3$ theory in AdS space, 
\begin{align}
\label{eq:twoptbubble}
    \mathcal{A}_{\text{2-bub}} = \int_{\AdS}\dd X_1 \dd X_2 E_{\Delta_1}(P_1,X_1)G_{\Delta}(X_1,X_2)G_{\Delta'}(X_1,X_2)E_{\Delta_2}(P_2,X_2) \ .
\end{align}
Its spectral representation, obtained as outlined in~\cref{sec:genstrat}, is~\cite{Giombi:2017hpr}
\begin{align}
\label{eq:twoptbubble_ITO_W}
    \mathcal{A}_{\text{2-bub}} = \frac{C_{\Delta_1}C_{\Delta_2}\Gamma\Big(\frac{\Delta_1+\Delta_2-d}{2}\Big)\,\mathcal{A}_{\text{2pt}}}{4\pi^h\Gamma(\Delta_1)\Gamma(\Delta_2)\Gamma\Big(d-\frac{\Delta_1+\Delta_2}{2}\Big)}\int\prod_{a=1}^{2}\left[\frac{\dd c_a}{2\pi i}\frac{1}{c_a-\delta_a}\right]\Wtwobubble\,,
\end{align}
where $\delta_1=\Delta-h$ and $\delta_2=\Delta'-h$ are the $h$-shifted dimensions of the internal fields and the potential $\Wtwobubble$ is
\begin{align}\label{eq:W2bub}
    \Wtwobubble &= \frac{1}{4c_1c_2\Gamma(c_1)\Gamma(-c_1)\Gamma(c_2)\Gamma(-c_2)}\\
    &\quad\times \textstyle{\Gamma\left(\frac{\Delta_1-c_1+c_2}{2}\right)\Gamma\left(\frac{\Delta_1+c_1+c_2}{2}\right)\Gamma\left(\frac{\Delta_2+c_1-c_2}{2}\right)\Gamma\left(\frac{\Delta_2-c_1-c_2}{2}\right)} \nonumber\\
    &\quad\times \textstyle{\Gamma\left(\frac{d-\Delta_1+c_1+c_2}{2}\right)\Gamma\left(\frac{d-\Delta_1-c_1+c_2}{2}\right)\Gamma\left(\frac{d-\Delta_2-c_1-c_2}{2}\right)\Gamma\left(\frac{d-\Delta_2+c_1-c_2}{2}\right)}\,.\nonumber
\end{align}
The evaluation of the factor $\mathcal{A}_{\text{2pt}}$ requires regularization\footnote{\label{ft:positionpart}As in the case of the tree-level two-point (boundary-boundary) two-point function, we do this 
by shifting the boundary dimension as $d\rightarrow d+\epsilon$, following~\cite{Giombi:2017hpr}. The result is
\begin{align*}
    \mathcal{A}_{\text{2pt}}(P_1,P_2) &= P_{12}^{\frac{d-\Delta_1-\Delta_2}{2}}\int_{\bAdS}\dd Q \frac{1}{(-2P_1\cdot Q)^{\frac{\Delta_1-\Delta_2+d}{2}}(-2P_2\cdot Q)^{\frac{\Delta_2-\Delta_1+d}{2}}}\nonumber\\
    &= \frac{2\pi^h}{\Gamma(h)} \frac{\delta_{\Delta_1\Delta_2}}{P_{12}^{\Delta_1}}\left[\frac{2}{\epsilon}+\log P_{12} + \log\pi -\psi(h)\right] \,,
\end{align*}
where $\psi$ is the digamma function.
The logarithmic term signals the relation between the two-point function and the anomalous dimension of the boundary operator.
}
and, as required by AdS symmetry, the result is proportional to $\delta_{\Delta_1\Delta_2}$. We will enforce the equality of the two boundary dimensions in the following and set $\Delta_1=\Delta_2 = \pmb{\Delta}$. This makes $\Wtwobubble$ odd in the two spectral parameters $c_1$ and $c_2$ and even under $c_1\leftrightarrow c_2$. 

We note that, in the limit in which any of the spectral parameters are large, the potential scales as
\begin{align}
    \frac{\Gamma(c_i/2)^4\Gamma(-c_i/2)^4}{\Gamma(c_i)\Gamma(-c_i)} \xrightarrow{c_i\rightarrow\infty} -\frac{16\pi^3}{c_i^3}\frac{\sin\pi c_i}{(\sin\frac{\pi c_i}{2})^4} \ ;
\end{align}
thus, the spectral parameter integrals can be computed via the residue theorem and, consequently, the arguments set forth in the previous section for the derivation of the SBP relations apply.  

Following our general strategy, we define the two-parameter family of integrals
\begin{align}
\label{eq:family2PTbubble}
    \Itwobubble{n_1}{n_2}(\delta_1,\delta_2,\pmb{\Delta},h) = \int\limits_{-i\infty}^{i\infty}\frac{\dd c_1}{2\pi i}\frac{\dd c_2}{2\pi i} \Pi_{n_1}(c_1-\delta_1)\Pi_{n_2}(c_2-\delta_2)\Wtwobubble(c_1,c_2,\pmb{\Delta},h)
\end{align}
from which $\Itwobubble{1}{1}(\delta_1,\delta_2,\pmb{\Delta},h)$ defines the two-point bubble amplitude \cref{eq:twoptbubble_ITO_W}, and begin with deriving the SBP relations. They rely on the properties of the potential under shifts of the spectral parameters. In our case they are
\begin{align}
    \frac{\Wtwobubble(c_1+2,c_2,\pmb{\Delta},h)}{\Wtwobubble(c_1,c_2,\pmb{\Delta},h)} = \prod_{\sigma=\pm 1}\frac{(\pmb{\Delta}+c_1+\sigma c_2)(2h-\pmb{\Delta}+c_1+\sigma c_2)}{(\pmb{\Delta}-c_1+\sigma c_2-2)(2h-\pmb{\Delta}-c_1+\sigma c_2-2)} \ .
\end{align}
The denominator of the right-hand side identifies the polynomial in \cref{initisbp}. It is not difficult to see that, with
\begin{align}
    P_{c_1,c_2}  = \prod_{\sigma=\pm 1}(\pmb{\Delta}-c_1+\sigma c_2)(2h-\pmb{\Delta}-c_1+\sigma c_2) \ ,
\end{align}
the product $P_{c_1+2,c_2} \Wtwobubble(c_1+2,c_2,\pmb{\Delta},h)$ does not exhibit a denominator and therefore such a shift maps the integrand in \cref{eq:family2PTbubble} to the same family. Starting from \cref{initisbp} with the $D_{c_1}^+$ difference operator it is straightforward albeit tedious to derive the general SBP relation
\begin{align}
\label{eq:2bubbleSBP}
    0 &= a_{n_1-3,n_2}\Itwobubble{n_1-3}{n_2} + a_{n_1-2,n_2}\Itwobubble{n_1-2}{n_2} + a_{n_1-1,n_2-2}\Itwobubble{n_1-1}{n_2-2} + a_{n_1-1,n_2-1}\Itwobubble{n_1-1}{n_2-1} \nonumber\\
    &\quad + a_{n_1-1,n_2}\Itwobubble{n_1-1}{n_2} + a_{n_1,n_2-2}\Itwobubble{n_1}{n_2-2} + a_{n_1,n_2-1}\Itwobubble{n_1}{n_2-1} + a_{n_1,n_2}\Itwobubble{n_1}{n_2} \nonumber\\
    &\quad + a_{n_1+1,n_2-4}\Itwobubble{n_1+1}{n_2-4} + a_{n_1+1,n_2-3}\Itwobubble{n_1+1}{n_2-3} + a_{n_1+1,n_2-2}\Itwobubble{n_1+1}{n_2-2} \nonumber\\
    &\quad + a_{n_1+1,n_2-1}\Itwobubble{n_1+1}{n_2-1} + a_{n_1+1,n_2}\Itwobubble{n_1+1}{n_2} \ ,
\end{align}
where the coefficients $a_{i,j}$ are given in \cref{eq:atwobubble} which we will use to derive difference equations in $\delta_{1,2}$ and $\pmb{\Delta}$. 
First however, we need to discuss boundary integrals.

To this end we make use of the observation similar to the one in the previous section that $n_i=-1$ represents the collapse of a bulk-bulk propagator. In our case, this reduces the bubble integral to a tadpole integral,
\begin{align}
    \int\limits_{-i\infty}^{i\infty}\frac{\dd c_1}{2\pi i}\Pi_{-1}(c_1-\delta_1)\Wtwobubble &=  \int\limits_{-i\infty}^{i\infty}\frac{\dd c_1}{2\pi i}\,c_1\,\Wtwobubble = \frac{\Gamma(d-\pmb{\Delta})\Gamma(\pmb{\Delta})\Gamma(h)^2}{\Gamma(2h)}\Wtadpole(c_2)\,,
\\
\label{eq:toTP}
    \Itwobubble{-1}{n_2} &= \frac{\Gamma(d-\pmb{\Delta})\Gamma(\pmb{\Delta})\Gamma(h)^2}{\Gamma(2h)} \Itadpole{n_2}(\delta_2)\,,
\\
    \Itwobubble{-2}{n_2} &= -2(\delta_1+3)\Itwobubble{-1}{n_2} \ ,
\end{align}
where in the last integral we used the oddness of $\Wtwobubble$ under $c_1\rightarrow -c_1$ to rewrite $\Pi_{-2}$ in terms of $\Pi_{-1}$. 
By setting $n_1=0$ in \cref{eq:2bubbleSBP}, it is not difficult to see that $\Itwobubble{-1}{n_2}$ and $\Itwobubble{-2}{n_2}$ also determine $\Itwobubble{-3}{n_2}$,
\begin{align}\label{eq:Im3n2}
    \Itwobubble{-3}{n_2} &= \Itwobubble{-1}{n_2-2} + 2(\delta_2-2n_2+3)\Itwobubble{-1}{n_2-1}\nonumber\\
    &\quad + \Big[\pmb{\Delta}(\pmb{\Delta}-2h)+(3\delta_1^2+24\delta_1+44)+(\delta_2-2n_2+2)^2\Big]\Itwobubble{-1}{n_2} \ ,
\end{align}
and iteratively all $\Itwobubble{-|n_1|}{n_2}$, in terms of tadpole integrals.\footnote{We note that \cref{eq:Im3n2} can also be derived without reliance on the SBP relation \eqref{eq:2bubbleSBP}. Indeed, by directly evaluating the $c_1$ integral, we find that
\begin{align*}
    \int\limits_{-i\infty}^{i\infty}\frac{\dd c_1}{2\pi i}\,c_1^3\,\Wtwobubble
    = \frac{\Gamma(d-\pmb{\Delta})\Gamma(\pmb{\Delta})\Gamma(h)^2}{\Gamma(2h)}\Big[c_2^2+\pmb{\Delta}(\pmb{\Delta}-2h)\Big]\Wtadpole(c_2) \ .
\end{align*}
Integrating this against $\Pi_{n_2}(c_2-\delta_2)$ and further using \cref{eq:toTP} yields
\cref{eq:Im3n2}. While this alternative strategy is not required here, it will be useful in more involved situations, such as the four-point bubble we discuss in \cref{sec:4bubble}.
}
Together with \cref{eq:zerotadpole} this further implies that 
\begin{align}
    \Itwobubble{n_1\leqslant 0}{n_2\leqslant 0} = 0 \ ,
\end{align}
which completes the set of boundary integrals. 


With the SBP relation(s) in hand, we now proceed to derive the desired difference equations in the arguments of $\Itwobubble{1}{1}(\delta_1, \delta_2, \pmb{\Delta})$. As before, we only write explicitly the relevant argument(s) for a given relation.
As $\Itwobubble{1}{1}(\delta_1, \delta_2, \pmb{\Delta})$ is symmetric under $\delta_1\leftrightarrow \delta_2$, so without loss of generality we focus on $\delta_1$. Since $\delta_i$ enter only in the argument of $\Pi_1$, it is easy to see using \cref{propertiesOfDPi} that
\begin{align}
    D_{\delta_1}^{-}\left[\Itwobubble{1}{1}\right] = - \Itwobubble{2}{1} \ .
\end{align}
To obtain the reduction of $\Itwobubble{2}{1}$, we set $n_1=n_2=1$ in \cref{eq:2bubbleSBP}. We will also need the reduction of $\Itwobubble{2}{-3}$, which can be obtained from \cref{eq:Im3n2} by setting $n_2=2$ and then exchanging $\delta_1$ and $\delta_2$. After some algebra, we find
\begin{align}
    &\Itwobubble{1}{1}(\delta_1-2) = \Itwobubble{1}{1}(\delta_1)\prod_{\sigma=\pm 1}\frac{(2h-\pmb{\Delta}-\delta_1+\sigma\delta_2)(\delta_1-\pmb{\Delta}+\sigma\delta_2)}{(2h-\pmb{\Delta}+\delta_1+\sigma\delta_2-2)(\delta_1+\pmb{\Delta}+\sigma\delta_2-2)} \nonumber\\
    \label{eq:InternalDiffEq}
    & - \frac{4\Gamma(2h-\pmb{\Delta})\Gamma(\pmb{\Delta})\Gamma(h)^2(2h-1)(\delta_1-1)\Itadpole{1}(\delta_2)}{\Gamma(2h)\prod_{\sigma=\pm 1}(2h-\pmb{\Delta}+\delta_1+\sigma\delta_2-2)(\delta_1+\pmb{\Delta}+\sigma\delta_2-2)} \\
    & - \frac{2\Gamma(2h-\pmb{\Delta})\Gamma(\pmb{\Delta})\Gamma(h)^2(2h-1)(\delta_1-1)(2 h^2 -2 h - 2 h \pmb{\Delta} + \pmb{\Delta}^2 + 
  2 \delta_1 - \delta_1^2 - \delta_2^2)\Itadpole{1}(\delta_1)}{\Gamma(2h)(\delta_1+h-1)(\delta_1+h-2)\prod_{\sigma=\pm 1}(2h-\pmb{\Delta}+\delta_1+\sigma\delta_2-2)(\delta_1+\pmb{\Delta}+\sigma\delta_2-2)} \ . \nonumber
\end{align}
In the next subsection, we will verify that the known expressions for the two-point bubble, as encoded by the anomalous dimension of the dual operator, satisfy this difference equation.

As discussed in \cref{sec:genstrat}, the derivation of a difference equation in $\pmb{\Delta}$ for $\Itwobubble{1}{1}$ cannot proceed directly by considering 
$D_{\pmb{\Delta}}^{+}[\Itwobubble{1}{1}]$ because shifts of $\pmb{\Delta}$ introduces new singularities which take us outside the $\Itwobubble{n_1}{n_2}$ family,
\begin{align}
\label{eq:WshiftDeltaBubble}
    \frac{\Wtwobubble(c_1,c_2,\pmb{\Delta}+2)}{\Wtwobubble(c_1,c_2,\pmb{\Delta})} = \prod_{\sigma_1,\sigma_2=\pm 1}\frac{\pmb{\Delta}+\sigma_1c_1+\sigma_2c_2}{2h-\pmb{\Delta}+\sigma_1c_1+\sigma_2c_2-2}
\end{align}
As discussed there, to solve this issue we introduce an auxiliary family of integrals, see eqs.~\eqref{WshiftsofDelta} and \eqref{auxiliaryfamily}.  
Introducing the polynomial
\begin{align}
    \widetilde{P}(c_1,c_2,\pmb{\Delta}) = \prod_{\sigma_1,\sigma_2=\pm 1}(2h-\pmb{\Delta}+\sigma_1c_1+\sigma_2c_2)\,,
\end{align}
referred to as $V$ in eqs.~\eqref{WshiftsofDelta} and \eqref{auxiliaryfamily}, and related to the denominator of the right-hand side of \cref{eq:WshiftDeltaBubble}, we consider the integral
\begin{align}\label{eq:Jbub}
    \mathcal{J} = \int\limits_{-i\infty}^{i\infty}\frac{\dd c_1}{2\pi i}\frac{\dd c_2}{2\pi i}\,\Pi_1(c_1-\delta_1)\Pi_2(c_2-\delta_2)\widetilde{P}(c_1,c_2,\pmb{\Delta})\Wtwobubble(c_1,c_2,\pmb{\Delta}) \ .
\end{align}
Similar to the polynomial entering the derivation of the SBP relations, $\widetilde{P}$ is chosen such that shifting $\pmb{\Delta}$ in
${\widetilde{P}}\times\Wtwobubble$ yields no poles beyond those already present in $\mathcal{J}$. We can evaluate the action of $D_{\pmb{\Delta}}^{+}$ on ${\cal J}$ in two different ways. On the one hand, we evaluate the action of this operator on the right-hand side of \cref{eq:Jbub}. On the other, we first expand ${\cal J}$ in terms of our basis of $\Itwobubble{n_1}{n_2}$ integrals and then act with $D_{\pmb{\Delta}}^{+}$ on the result.
The former path leads to
\begin{align}\label{eq:DJ1}
    D_{\pmb{\Delta}}^{+}\big[\mathcal{J}\big] &= \int\limits_{-i\infty}^{i\infty}\frac{\dd c_1}{2\pi i}\frac{\dd c_2}{2\pi i}\,\Pi_1(c_1-\delta_2)\Pi_2(c_2-\delta_2)D_{\pmb{\Delta}}^{+}\Big[\widetilde{P}(c_1,c_2,\pmb{\Delta})\Wtwobubble(c_1,c_2,\pmb{\Delta})\Big]\nonumber\\
    &= 4h(\pmb{\Delta}-h)(2 h^2 - 2 h \pmb{\Delta} + \pmb{\Delta}^2 - \delta_1^2 - \delta_2^2) \Itwobubble{1}{1}(\pmb{\Delta}) \nonumber\\
    &\quad - \frac{4h(\pmb{\Delta}-h)\Gamma(2h-\pmb{\Delta})\Gamma(\pmb{\Delta})\Gamma(h)^2}{\Gamma(2h)}\Big[\Itadpole{1}(\delta_1)+\Itadpole{1}(\delta_2)\Big] \ .
\end{align}
The latter first organizes ${\cal J}$ as
\begin{align}\label{eq:Jbub2}
    \mathcal{J} = \widetilde{P}(\delta_1,\delta_2,\pmb{\Delta})\Itwobubble{1}{1}(\pmb{\Delta}) & + \frac{\Gamma(2h-\pmb{\Delta})\Gamma(\pmb{\Delta})\Gamma(h)^2}{\Gamma(2h)}\Big[(6h\pmb{\Delta}-8h^2+\pmb{\Delta}^2-\delta_1^2+\delta_2^2)\Itadpole{1}(\delta_1) \nonumber\\
    &\qquad\qquad\qquad +(6h\pmb{\Delta}-8h^2+\pmb{\Delta}^2+\delta_1^2-\delta_2^2)\Itadpole{1}(\delta_2)\Big] \ .
\end{align}
All $\pmb{\Delta}$ dependence is manifest except for the desired integral $\Itwobubble{1}{1}(\pmb{\Delta})$. Thus, acting with $D_{\pmb{\Delta}}^{+}$ on \cref{eq:Jbub2} and equating the result with \cref{eq:DJ1} yields, after a slight reorganization, 
\begin{align}
    &\Itwobubble{1}{1}(\pmb{\Delta}+2) = \Itwobubble{1}{1}(\pmb{\Delta})\prod_{\sigma_1,\sigma_2=\pm 1}\frac{\pmb{\Delta}+\sigma_1\delta_1+\sigma_2\delta_2}{\pmb{\Delta}-2h+2+\sigma_1\delta_1+\sigma_2\delta_2} \nonumber \\
    \label{eq:DelDiffEq}
    &\quad - \frac{2(2h-1)(\pmb{\Delta}-h+1)\Gamma(2h-\pmb{\Delta}-2)\Gamma(\pmb{\Delta})\Gamma(h)^2}{\Gamma(2h)\prod_{\sigma_1,\sigma_2=\pm 1}(\pmb{\Delta}-2h+2+\sigma_1\delta_1+\sigma_2\delta_2)} \\
    &\quad\quad\times\Big[(\pmb{\Delta}^2-2h\pmb{\Delta}+2\pmb{\Delta}+\delta_1^2-\delta_2^2)\Itadpole{1}(\delta_2) +(\pmb{\Delta}^2-2h\pmb{\Delta}+2\pmb{\Delta}+\delta_2^2-\delta_1^2)\Itadpole{1}(\delta_1)\Big] \nonumber \ ,
\end{align}
which is the desired difference equation in $\pmb{\Delta}$.

To summarize, \cref{eq:InternalDiffEq,eq:DelDiffEq} are the difference equations determining (or at least constraining) the two-point one-loop bubble integral $\Itwobubble{1}{1}$, which determines the two-point one-loop AdS correlator \eqref{eq:twoptbubble} and \eqref{eq:twoptbubble_ITO_W}. 
It is tempting to speculate that the physical conditions uniquely specifying the solutions of these equations may include:
\begin{itemize}
\item the two-point function, and hence $\Itwobubble{1}{1}$, should be real
\item all poles should be simple
\item a finite number of poles in $\delta_i>0$ for any fixed value of $\pmb{\Delta}$ (in analogy with the flat space case in which a finite number of particles can be produced for fixed incoming energy)
\item all poles in $\delta_i$ for sufficiently large $\pmb{\Delta}$ should have residues of the same sign (in analogy with the flat space case in which the sign of the residue indicates whether or not a pole corresponds to physical states)
\item we may expect that the poles are spaced by one unit and correspond to conformal descendants. 
\end{itemize}
We will however not attempt to solve these equations and fully identify the physical constraints that render the solution unique.
Instead, we will verify that the known results solve these equations. 

\subsection{Analytic results for two-point bubble}\label{sec:2bubblecomparison}

The two-point bubble diagram was discussed at length in  Ref.~\cite{Giombi:2017hpr}, where the information not constrained by AdS (conformal) symmetry is captured by the (contribution to the) anomalous dimension $\gamma^{\text{2-bub}}$ of the dual boundary operator. In our notation, this anomalous dimension (which is the coefficient of $\log P_{12}$ in \cref{eq:twoptbubble_ITO_W} with 
$\mathcal{A}_{\text{2pt}}(P_1,P_2)$ given in footnote~\ref{ft:positionpart}) is
\begin{align}
    \gamma^{\text{2-bub}}(\delta_1,\delta_2,\pmb{\Delta},h) = -\frac{\Itwobubble{1}{1}(\delta_1,\delta_2,\pmb{\Delta},h)}{2\pi^h\Gamma(h)\Gamma(2h-\pmb{\Delta})\Gamma(\pmb{\Delta})(\pmb{\Delta}-h)} \ .
\end{align}
It is not difficult, albeit slightly tedious, to reorganize eqs.~\eqref{eq:InternalDiffEq} and \eqref{eq:DelDiffEq} as difference equations for $\gamma^{\text{2-bub}}(\delta_1,\delta_2,\pmb{\Delta},h)$. For the dimension of the internal state, the result is
\begin{align}\label{eq:DEbub2}
    &\gamma^{\text{2-bub}}(\delta_1-2) = \gamma^{\text{2-bub}}(\delta_1)\prod_{\sigma=\pm 1}\frac{(2h-\pmb{\Delta}-\delta_1+\sigma\delta_2)(\delta_1-\pmb{\Delta}+\sigma\delta_2)}{(2h-\pmb{\Delta}+\delta_1+\sigma\delta_2-2)(\delta_1+\pmb{\Delta}+\sigma\delta_2-2)} \nonumber\\
    &\quad - \frac{4(2h-1)(\delta_1-1)\mathcal{G}^{\text{tad}}(\delta_2,h)}{(\pmb{\Delta}-h)\prod_{\sigma=\pm 1}(2h-\pmb{\Delta}+\delta_1+\sigma\delta_2-2)(\delta_1+\pmb{\Delta}+\sigma\delta_2-2)} \\
    &\quad -  \frac{2(2h-1)(\delta_1-1)(2 h^2 -2 h - 2 h \pmb{\Delta} + \pmb{\Delta}^2 + 
  2 \delta_1 - \delta_1^2 - \delta_2^2)\mathcal{G}^{\text{tad}}(\delta_1,h)}{(\pmb{\Delta}-h)(\delta_1+h-1)(\delta_1+h-2)\prod_{\sigma=\pm 1}(2h-\pmb{\Delta}+\delta_1+\sigma\delta_2-2)(\delta_1+\pmb{\Delta}+\sigma\delta_2-2)}\,, \nonumber
\end{align}
where $\mathcal{G}^{\text{tad}}$ is given in \cref{eq:tadpoleresult}. 
Similarly, reorganizing \cref{eq:DelDiffEq} into a difference equation for $\gamma^{\text{2-bub}}$ leads to
\begin{align}\label{eq:DEbubExt}
    &\gamma^{\text{2-bub}}(\pmb{\Delta}+2) \nonumber\\
    &= \frac{\gamma^{\text{2-bub}}(\pmb{\Delta})(2h-\pmb{\Delta}-2)(2h-\pmb{\Delta}-1)(\pmb{\Delta}-h)}{\pmb{\Delta}(\pmb{\Delta}+1)(\pmb{\Delta}-h+2)}\prod_{\sigma_1,\sigma_2=\pm 1}\frac{\pmb{\Delta}+\sigma_1\delta_1+\sigma_2\delta_2}{\pmb{\Delta}-2h+2+\sigma_1\delta_1+\sigma_2\delta_2} \nonumber\\
    &\quad - \frac{2(2h-1)(\pmb{\Delta}-h+1)}{\pmb{\Delta}(\pmb{\Delta}+1)(\pmb{\Delta}-h+2)\prod_{\sigma_1,\sigma_2=\pm 1}(\pmb{\Delta}-2h+2+\sigma_1\delta_1+\sigma_2\delta_2)} \\
    &\quad\quad\times\Big[(\pmb{\Delta}^2-2h\pmb{\Delta}+2\pmb{\Delta}+\delta_1^2-\delta_2^2)\mathcal{G}^{\text{tad}}(\delta_2,h) +(\pmb{\Delta}^2-2h\pmb{\Delta}+2\pmb{\Delta}+\delta_2^2-\delta_1^2)\mathcal{G}^{\text{tad}}(\delta_1,h)\Big] \ , \nonumber
\end{align}
where as before $\mathcal{G}^{\text{tad}}$ is given in \cref{eq:tadpoleresult}. 
%

Now we consider the known analytic expressions for the anomalous dimension. First, for $h=1$ but arbitrary bulk and boundary conformal weights, Ref.~\cite{Giombi:2017hpr} provides an analytic expression for this anomalous dimension, $\gamma^{\text{2-bub}}(\delta_1,\delta_2,\pmb{\Delta},h=1)$, see eq.~(2.47) there. A simpler but equivalent form is  
\begin{align}\label{eq:bub3d}
    \gamma^{\text{2-bub}}(\delta_1,\delta_2,\pmb{\Delta},h=1) = \frac{1}{8\pi(\pmb{\Delta}-1)^2}\left(H_{\frac{\delta_1+\delta_2-\pmb{\Delta}}{2}}-H_{\frac{\delta_1+\delta_2+\pmb{\Delta}-2}{2}}\right) \ .
\end{align}
We have checked that indeed this expression obeys both \cref{eq:DEbub2,eq:DEbubExt}.

Next, we consider the case in which the internal conformal weights are the same while the space-time dimension is arbitrary. For $\delta_1=\delta_2=\Delta-h$, we define for convenience
\begin{align}
    \tilde{\gamma}^{\text{2-bub}}({\Delta},{\Delta},\widetilde{{\Delta}},h) \equiv \gamma^{\text{2-bub}}({\Delta}-h,{\Delta}-h,\widetilde{{\Delta}},h)\,,
\end{align}
where we also relabel the boundary weight as $\pmb{\Delta}\rightarrow\widetilde{\Delta}$.
This anomalous dimension also has an analytic expression~\cite{Carmi:2018qzm},
\begin{align}\label{eq:2bub_special}
    &\tilde{\gamma}^{\text{2-bub}}({\Delta},{\Delta},\widetilde{{\Delta}},h) = -(\widetilde{{\Delta}}-h)^{-1}\times\frac{\Gamma({\Delta})\Gamma({\Delta}-h+1/2)\Gamma(2{\Delta}-h)}{4(4\pi)^h}\\
    &\qquad\times\left(\Gamma\left[\frac{2{\Delta}-\widetilde{{\Delta}}}{2}\right]{}_{5}\tilde{F}_{4}\left[\begin{array}{c}
    h\,,\,{\Delta}\,,\,{\Delta}-h+\frac{1}{2}\,,\,\frac{2{\Delta}-\widetilde{{\Delta}}}{2}\,,\,2{\Delta}-h \\
    {\Delta}+\frac{1}{2}\,,\,{\Delta}-h+1\,,\,\frac{2{\Delta}-\widetilde{{\Delta}}}{2}+1\,,\,2{\Delta}-2h+1
    \end{array};1\right]\right. \nonumber\\
    &\qquad\quad\left. +\;\Gamma\left[\frac{2{\Delta}+\widetilde{{\Delta}}}{2}-h\right]{}_5\tilde{F}_4\left[\begin{array}{c}
    h\,,\,{\Delta}\,,\,{\Delta}-h+\frac{1}{2}\,,\,\frac{2{\Delta}+\widetilde{{\Delta}}}{2}-h\,,\,2{\Delta}-h \\
    {\Delta}+\frac{1}{2}\,,\,{\Delta}-h+1\,,\,\frac{2{\Delta}+\widetilde{{\Delta}}}{2}-h+1\,,\,2{\Delta}-2h+1
    \end{array};1\right]\right) \ ,\nonumber
\end{align}
where ${}_5\tilde{F}_4$ is the regularized hypergeometric function, which is related to the usual hypergeometric function by
\begin{align}
{}_{p}\tilde{F}_{q}\left[\begin{array}{c}
    a_1,\dots a_p\\
    b_1,\dots, b_q
    \end{array};z\right]
    =\frac{1}{\Gamma(b_1)\dots \Gamma(b_q)}
{}_{p}{F}_{q}\left[\begin{array}{c}
    a_1,\dots a_p\\
    b_1,\dots, b_q
    \end{array};z\right] \ .
\end{align}
The anomalous dimension is obtained through a relation to the corresponding spectral bubble,
\begin{align}
    \tilde{\gamma}^{\text{2-bub}}({\Delta},{\Delta},\widetilde{{\Delta}},h) = - \frac{\tilde{B}(\nu)}{i\nu}\bigg|_{i\nu = \widetilde{{\Delta}}-h}\,,
\end{align}
where $\tilde{B}(\nu)$ is given by eq.~(4.27) of Ref.~\cite{Carmi:2018qzm}. We have verified that $\tilde{\gamma}^{\text{2-bub}}({\Delta},{\Delta},\widetilde{{\Delta}},h)$ indeed satisfies the difference equation in the boundary weight,
\begin{align}
    &\tilde{\gamma}^{\text{2-bub}}({\Delta},{\Delta},\widetilde{{\Delta}}+2,h) \nonumber\\
    &= - \frac{\widetilde{{\Delta}}(\widetilde{{\Delta}}-2{\Delta}+2h)(\widetilde{{\Delta}}+2{\Delta}-2h)(2h-\widetilde{{\Delta}}-1)(\widetilde{{\Delta}}-h) \, \tilde{\gamma}^{\text{2-bub}}({\Delta},{\Delta},\widetilde{{\Delta}},h)}{(\widetilde{{\Delta}}-2h+2)(\widetilde{{\Delta}}-2{\Delta}+2)(\widetilde{{\Delta}}+2{\Delta}-4h+2)(\widetilde{{\Delta}}+1)(\widetilde{{\Delta}}-h+2)} \nonumber\\
    & \quad - \frac{4(2h-1)(\widetilde{{\Delta}}-h+1)\,\mathcal{G}^{\text{tad}}({\Delta}-h,h)}{(\widetilde{{\Delta}}+1)(\widetilde{{\Delta}}-h+2)(\widetilde{{\Delta}}-2h+2)(\widetilde{{\Delta}}-2{\Delta}+2)(\widetilde{{\Delta}}+2{\Delta}-4h+2)}\,,
    \label{eq:2ptgammaspecial}
\end{align}
which is derived from \cref{eq:DelDiffEq} by sending $\pmb{\Delta}\rightarrow\widetilde{\Delta}$ and $\delta_1=\delta_2=\Delta-h$. We therefore see that the difference equations \eqref{eq:InternalDiffEq} and \eqref{eq:DelDiffEq} are satisfied by the known analytic expressions for two-point bubble integrals. They provide a consistency check for future direct evaluations of more general cases.

Interestingly, we can now use the difference equation~\eqref{eq:DEbub2} to derive new analytic expressions for the anomalous dimensions (and the two-point bubble integrals) with the two internal conformal weights differ by an even integer. Plugging \cref{eq:2bub_special} into the right hand side of \cref{eq:DEbub2}, we get
\begin{align}\label{eq:2bub_sp2}
    &\tilde{\gamma}^{\text{2-bub}}(\Delta-2,\Delta,\widetilde{\Delta},h) \nonumber\\
    &\quad = \frac{\mathcal{Y}}{\widetilde{\Delta}-h}+\frac{\mathcal{X}_1}{\widetilde{\Delta}-h}{}_{5}\tilde{F}_{4}\left[\begin{array}{c}
    h\,,\,{\Delta}\,,\,{\Delta}-h+\frac{1}{2}\,,\,\frac{2{\Delta}-\widetilde{{\Delta}}}{2}\,,\,2{\Delta}-h \\
    {\Delta}+\frac{1}{2}\,,\,{\Delta}-h+1\,,\,\frac{2{\Delta}-\widetilde{{\Delta}}}{2}+1\,,\,2{\Delta}-2h+1
    \end{array};1\right] \nonumber\\
    & \quad\quad + \frac{\mathcal{X}_2}{\widetilde{\Delta}-h} {}_5\tilde{F}_4\left[\begin{array}{c}
    h\,,\,{\Delta}\,,\,{\Delta}-h+\frac{1}{2}\,,\,\frac{2{\Delta}+\widetilde{{\Delta}}}{2}-h\,,\,2{\Delta}-h \\
    {\Delta}+\frac{1}{2}\,,\,{\Delta}-h+1\,,\,\frac{2{\Delta}+\widetilde{{\Delta}}}{2}-h+1\,,\,2{\Delta}-2h+1
    \end{array};1\right],
\end{align}
where the coefficients are 
\begin{align}
    \mathcal{X}_1 &= \frac{\widetilde{\Delta}(\widetilde{\Delta}-2h)(\widetilde{\Delta}+2\Delta-4h)(\widetilde{\Delta}-2\Delta+2h)\Gamma(\Delta)\Gamma(\Delta{-}h{+}\frac{1}{2})\Gamma(\Delta{-}\frac{\widetilde{\Delta}}{2}{-}1)\Gamma(2\Delta{-}h)}{8(4\pi)^h(\widetilde{\Delta}-2)(\widetilde{\Delta}-2h+2)(\widetilde{\Delta}+2\Delta-2h-2)}\,,\nonumber\\
    \mathcal{X}_2 &= \mathcal{X}_1\Big|_{\widetilde{\Delta}\rightarrow-\widetilde{\Delta}+2h}\,,\\
    \mathcal{Y} &= \frac{(2h-1)(\Delta-h-1)(4\Delta-4h\Delta-\widetilde{\Delta}^2+2h\widetilde{\Delta}+4h-4)\,(\Delta-2h+1)_{2h-3}}{2\pi^h\cos(\pi h)(\widetilde{\Delta}-2)(\widetilde{\Delta}-2h+2)(\widetilde{\Delta}+2\Delta-2h-2)(\widetilde{\Delta}-2\Delta+2)\,(h)_{h}}\,. \nonumber
\end{align}
Iterating this process will lead to an expression for $\tilde{\gamma}^{\text{2-bub}}(\Delta-2n,\Delta,\widetilde{\Delta},h)$.

Exchanging $\delta_1$ with $\delta_2$ in \cref{eq:DEbub2} gives a difference equation in $\delta_2$, which can be used to derive $\gamma^{\text{2-bub}}(\Delta-2,\Delta-2,\widetilde{\Delta},h)$ from $\gamma^{\text{2-bub}}(\Delta-2,\Delta,\widetilde{\Delta},h)$ given in \cref{eq:2bub_sp2}. As another consistency check, we have verified that $\gamma^{\text{2-bub}}(\Delta-2,\Delta-2,\widetilde{\Delta},h)$ obtained through our difference equations agrees exactly with \cref{eq:2bub_special} under $\Delta\rightarrow\Delta-2$.

We note that, similar to the case of the tree-level scalar exchange discussed at the end of \cref{sec:scalar}, this difference equation implies that, for special values of the boundary conformal dimension $\widetilde{{\Delta}}$ which we will denote as $\widetilde{{\Delta}}_*$, the anomalous dimension $\tilde{\gamma}^{\text{2-bub}}$ becomes a simple ratio of Gamma functions. Indeed, for $\widetilde{{\Delta}}_* = 2h-1$ and $\widetilde{\Delta}_{*}=2\Delta-2h$, the coefficient of $\tilde{\gamma}^{\text{2-bub}}({\Delta},{\Delta},\widetilde{{\Delta}},h)$
on the right-hand side of \cref{eq:2ptgammaspecial} vanishes and we are left only the inhomogeneous term:
\begin{align}\label{eq:2ptgammaspecial_very}
    &\tilde{\gamma}^{\text{2-bub}}({\Delta},{\Delta},2h+1,h) = - \frac{2(2h-1)\,\mathcal{G}^{\text{tad}}({\Delta}-h,h)}{(h+1)(2h-2{\Delta}+1)(2{\Delta}-2h+1)}\,, \\
    &\tilde{\gamma}^{\text{2-bub}}(\Delta,\Delta,2\Delta-2h+2,h) = \frac{(2h-1)\,\mathcal{G}^{\text{tad}}({\Delta}-h,h)}{2(h-1)(2\Delta-2h+1)(2\Delta-3h+2)(\Delta-2h+1)}\,,
\end{align}
with $\tilde{\gamma}^{\text{2-bub}}$ given by \cref{eq:2bub_special}. Note that $\widetilde{\Delta}_{*}=h$ is not a zero for the first term of \cref{eq:2ptgammaspecial} because it is also a pole for $\tilde{\gamma}^{\text{2-bub}}({\Delta},{\Delta},\widetilde{{\Delta}},h)$ following \cref{eq:2bub_special}.\footnote{Note that the origin of this divergence is from the Gamma function factor in the correlator, see \cref{eq:twoptbubble_ITO_W}. The integral $\mathcal{I}^{\text{2-bub}}$ remains to be finite.}
The other zero of the coefficient of $\tilde{\gamma}^{\text{2-bub}}({\Delta},{\Delta},\widetilde{{\Delta}},h)$
on the right-hand side of \cref{eq:2ptgammaspecial}, $\widetilde{\Delta}_{*}=2h-2\Delta$, is at an unphysical 
value for $\widetilde{{\Delta}}$.
We have verified that the expression in \cref{eq:2bub_special} reproduces these simple expressions.

Returning to \cref{eq:DEbub2,eq:DEbubExt} and using this observation we see that, for $\pmb{\Delta} \in \{\delta_1\pm \delta_2, 2h- \delta_1\pm\delta_2\}$ and $\pmb{\Delta} \in \{2h-1, 2h-2, \delta_1+\delta_2\}$, respectively, and assuming that ${\gamma}^{\text{2-bub}}$ has no poles at these locations, the difference equations determines the anomalous dimension as given by the the inhomogeneous term, which is a rational combination of Gamma functions.

Finally, we note that the two-point bubble has a generic expression in terms of a ${}_{9}F_{8}$ function,
\begin{align}
    \gamma^{\text{2-bub}}(\delta_1,\delta_2,\pmb{\Delta},h) = \frac{-1}{\pmb{\Delta}-h} B(\lambda,\nu,\kappa)\,,
\end{align}
where $B(\lambda,\nu,\kappa)$ is given in Eq.~(A.29) of~\cite{Cacciatori:2024zbe}. To match our notation, one needs to pick $\lambda=\delta_1$, $\nu=\delta_2$, $\kappa=-i(\pmb{\Delta}-h)$, $\delta_{\text{there}}=h$ and $d_{\text{there}}=d+1$ in Eq.~(A.29) of~\cite{Cacciatori:2024zbe}. We have checked that this result indeed satisfies our difference equation~\eqref{eq:DEbub2} and~\eqref{eq:DEbubExt}.

\subsection{Three-point bubble}\label{subsec:bub3}

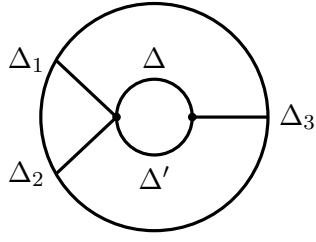
\begin{figure}
    \centering
    \begin{tikzpicture}[every path/.style={very thick}]
        \pgfmathsetmacro{\r}{1.5};
        \draw (0,0) circle (\r);
        \draw (0,0) circle (\r/3);
        \node at (0,\r/3) [above=0pt] {$\Delta$};
        \node at (0,-\r/3) [below=0pt] {$\Delta'$};
        \filldraw (\r/3,0) circle (1pt) (-\r/3,0) circle (1pt);
        \draw (\r/3,0) -- (\r,0) node [right=0pt] {$\Delta_3$};
        \draw (-\r/3,0) -- (150:\r) node [left=0pt] {$\Delta_1$};
        \draw (-\r/3,0) -- (-150:\r) node [left=0pt] {$\Delta_2$};
    \end{tikzpicture}
    \caption{The three-point bubble.}
    \label{fig:3bub}
\end{figure}

Let us proceed to the integral next in complexity, the three-point bubble integral in $\phi_1\phi_2\phi_3\oplus \phi_2\phi_3\phi_4\phi_5$ scalar theory in AdS space. As shown in \cref{fig:3bub}, it is defined as
\begin{equation}
\mathcal{A}_{\text{3-bub}}=\int_{\AdS} \dd X_{1}\dd X_{2}\, E_{\Delta_{3}}(P_{3},X_{1})\,G_{\Delta}(X_{1},X_{2})\,G_{\Delta'}(X_{1},X_{2})\,E_{\Delta_{2}}(P_{2},X_{2})\,E_{\Delta_{1}}(P_{1},X_{2})\, .
\end{equation}
While, as discussed in \cref{sec:2bubblecomparison}, in a gauge/gravity duality framework the two-point correlation function corrects the anomalous dimension of the boundary operator corresponding to the AdS field, the three-point bubble as defined here represents one of the several contributions to the OPE coefficient.\footnote{Other contributions are the triangle integral, as well as possible leftovers after the anomalous dimension was extracted from the two-point correlation function attached to a three-point vertex. }

The construction of the potential for the three-point bubble integral follows the same steps as in the case of the two-point bubble integral. In the spectral representation, the three-point bubble is given by 
\begin{align}
    \mathcal{M}_{\text{3-bub}} 
    &= \frac{1}{2\pi^h\Gamma(h)\Gamma(2h-\Delta_3)\Gamma(\Delta_3)}\int\prod_{i=1}^{2}\left[\frac{\dd c_i}{2\pi i}\frac{1}{c_i-\delta_i}\right]\Wthreebub
    \\
    &\equiv
    \frac{{\cal I}^\text{3-bub}_{1,1}}{2\pi^h\Gamma(h)\Gamma(2h-\Delta_3)\Gamma(\Delta_3)}
\end{align}
where, as in \cref{sec:bub2}, $\delta_1=\Delta-h$ and $\delta_2=\Delta'-h$ are the shifted internal dimensions. The potential~$\Wthreebub$ is
\begin{align}
\label{eq:W3bub}
    \Wthreebub &= \frac{1}{4c_1c_2\Gamma(c_1)\Gamma(-c_1)\Gamma(c_2)\Gamma(-c_2)}\\
    &\quad\times \textstyle{\Gamma\left(\frac{\Delta_3-c_1+c_2}{2}\right)\Gamma\left(\frac{\Delta_3+c_1+c_2}{2}\right)\Gamma\left(\frac{\Delta_3+c_1-c_2}{2}\right)\Gamma\left(\frac{\Delta_3-c_1-c_2}{2}\right)} \nonumber\\
    &\quad\times \textstyle{\Gamma\left(\frac{d-\Delta_3+c_1+c_2}{2}\right)\Gamma\left(\frac{d-\Delta_3-c_1+c_2}{2}\right)\Gamma\left(\frac{d-\Delta_3-c_1-c_2}{2}\right)\Gamma\left(\frac{d-\Delta_3+c_1-c_2}{2}\right)}\,.\nonumber
\end{align}
It is easy to see by directly comparing \cref{eq:W3bub} and  \cref{eq:W2bub} that 
\begin{align}
\Wthreebub = \Wtwobubble\Big|_{\Delta_1=\Delta_2\rightarrow\Delta_3} \ .
\end{align}
Therefore, the corresponding difference equations~\eqref{eq:InternalDiffEq} and \eqref{eq:DelDiffEq} and SBP relations~\eqref{eq:2bubbleSBP} derived for two-point bubbles in \cref{sec:bub2} extend, with the identification $\pmb{\Delta}\rightarrow \Delta_3$, change to the three-point bubble integral ${\cal I}^\text{3-bub}_{1,1}$.

\subsection{Four-point bubble}\label{sec:4bubble}

\begin{figure}
    \centering
    \subfloat[]{\label{fig:4buba}
    \begin{tikzpicture}[every path/.style={very thick}]
        \pgfmathsetmacro{\r}{1.5};
        \draw (0,0) circle (\r);
        \draw (0,0) circle (\r/3);
        \node at (0,\r/3) [above=0pt] {$\Delta$};
        \node at (0,-\r/3) [below=0pt] {$\Delta'$};
        \filldraw (\r/3,0) circle (1pt) (-\r/3,0) circle (1pt);
        \draw (\r/3,0) -- (30:\r) node [right=0pt] {$\Delta_3$};
        \draw (\r/3,0) -- (-30:\r) node [right=0pt] {$\Delta_4$};
        \draw (-\r/3,0) -- (150:\r) node [left=0pt] {$\Delta_1$};
        \draw (-\r/3,0) -- (-150:\r) node [left=0pt] {$\Delta_2$};
    \end{tikzpicture}}
    \qquad
    \subfloat[]{\label{fig:4bubb}
    \begin{tikzpicture}[every path/.style={very thick}]
        \pgfmathsetmacro{\r}{1.5};
        \pgfmathsetmacro{\s}{\r/6};
        \draw (0,0) circle (\r);
        \draw (\s,0) circle (\r/3);
        \coordinate (l) at (\s-\r/3,0);
        \coordinate (r) at (\s+\r/3,0);
        \coordinate (l2) at (\s-2*\r/3,0);
        \node at (\s,\r/3) [above=0pt] {$\Delta$};
        \node at (\s,-\r/3) [below=0pt] {$\Delta'$};
        \filldraw (l) circle (1pt) (r) circle (1pt) (l2) circle (1pt);
        \draw (r) -- (30:\r) node [right=0pt] {$\Delta_3$};
        \draw (r) -- (-30:\r) node [right=0pt] {$\Delta_4$};
        \draw (l2) -- (150:\r) node [left=0pt] {$\Delta_1$};
        \draw (l2) -- (-150:\r) node [left=0pt] {$\Delta_2$};
        \draw (l2) -- (l) node [above=0pt,pos=0.5] {$\widetilde{\Delta}$};
    \end{tikzpicture}}
    \caption{(a) The four-point bubble in $\phi^4$ theory and (b) the four-point bubble with an auxiliary bulk-bulk propagator.}
    \label{fig:4bub}
\end{figure}
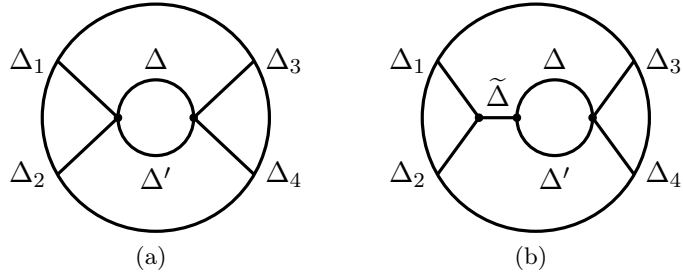

Finally, we consider the the four-point bubble in the $\phi_1\phi_2\phi_3\phi_4\oplus\phi_3\phi_4\phi_5\phi_6$ theory, as shown in \cref{fig:4buba},
\begin{align}\label{eq:4bub}
    \mathcal{A}_{\text{4-bub}} = \int_{\AdS} \dd X_1 \dd X_2 \, & E_{\Delta_1}(P_1,X_1) E_{\Delta_2}(P_2,X_1) G_{\Delta}(X_1,X_2) G_{\Delta'}(X_1,X_2) \nonumber\\
    & \times E_{\Delta_3}(P_3,X_2) E_{\Delta_4}(P_4,X_2) \,.
\end{align}
The corresponding Mellin amplitude is obtained by following the strategy in \cref{sec:genstrat}. However, the potential function for this diagram is a generalized hypergeometric function, which is inconvenient for deriving SBP relations and difference equations. Instead, we insert an auxiliary bulk-bulk propagator as in \cref{fig:4bubb} and consider 
\begin{align}\label{eq:4bubaux}
    \widetilde{\mathcal{A}}_{\text{4-bub}} = \int_{\AdS}\dd X_1 \dd X_2 \dd X_3 \, & E_{\Delta_1}(P_1,X_1)E_{\Delta_2}(P_2,X_1)G_{\widetilde{\Delta}}(X_1,X_3)G_{\Delta}(X_3,X_2) \nonumber\\
    &\times G_{\Delta'}(X_3,X_2)E_{\Delta_3}(P_3,X_2)E_{\Delta_4}(P_4,X_2)\,.
\end{align}
The additional propagator makes the potential function $\Wfourbub$ a product of gamma functions,
\begin{align}
    \Wfourbub(c,c_1,c_2,s,\Delta_{12},\Delta_{34}) &= \frac{1}{32 \, c \, c_1 c_2} \prod_{\sigma=\pm 1}\frac{\Gamma(\frac{h-s+\sigma c}{2})\Gamma(\frac{\Delta_{12}-h+\sigma c}{2})\Gamma(\frac{\Delta_{34}-h+\sigma c}{2})}{\Gamma(\sigma c)\Gamma(h+\sigma c)\Gamma(\sigma c_1)\Gamma(\sigma c_2)}\nonumber\\
    &\quad\times\prod_{\sigma,\sigma_1,\sigma_2=\pm 1}\Gamma\left(\frac{h+\sigma c+\sigma_1 c_1+\sigma_2 c_2}{2}\right)\, ,
\end{align}
which is in the general class of potentials considered in \cref{sec:genstrat}.

Following the strategy outlined in \cref{sec:genstrat} and similarly to the previous examples, we define the family of integrals 
\begin{gather}
    \Ifourbub{n}{n_1}{n_2} = \int\limits_{-i\infty}^{i\infty}\frac{\dd c}{2\pi i}\frac{\dd c_1}{2\pi i}\frac{\dd c_2}{2\pi i}\Pi_{n}(c-\delta)\Pi_{n_1}(c_1-\delta_1)\Pi_{n_2}(c_2-\delta_2)\Wfourbub(c,c_1,c_2,s) \\
    \delta\equiv\widetilde{\Delta}-h\,,\qquad \delta_1 = \Delta-h\,,\qquad \delta_2\equiv\Delta'-h\, ; \nonumber
\end{gather}
the Mellin amplitude for \cref{eq:4bub} is then
\begin{align}\label{eq:M4bub_ME}
    \mathcal{M}_{\text{4-bub}} = \frac{\Ifourbub{-1}{1}{1}}{\pi^h \Gamma(h)\Gamma\left(\frac{\Delta_{1234}}{2}-h\right)\Gamma\left(\frac{\Delta_{12}-s}{2}\right)\Gamma\left(\frac{\Delta_{34}-s}{2}\right)}\,.
\end{align}
Recall that, as discussed in previous sections, an index $-1$ corresponds to the collapse of a bulk-bulk propagator, so in $\Ifourbub{-1}{1}{1}$ the auxiliary propagator is collapsed. We note that 
\begin{align}
0 = \Ifourbub{n}{0}{n_2} = \Ifourbub{n}{n_1}{0} = \Ifourbub{0}{n_1}{n_2} \ ,
\label{eq:4bub_zeroes}
\end{align}
because $\Wfourbub$ is odd in $c$, $c_1$ and $c_2$. Moreover, because
\begin{align}
    0 = \cint{c}\,c^{2k}\Wfourbub = \cint{c_1}\,c_1^{2k}\Wfourbub = \cint{c_2}\,c_2^{2k}\Wfourbub\,,
\end{align}
we can write $\Ifourbub{-2k}{n_1}{n_2}$, $\Ifourbub{n}{-2k}{n_2}$ and $\Ifourbub{n}{n_1}{-2k} $ can be written in terms of $\Ifourbub{-2l+1}{n_1}{n_2}$, $\Ifourbub{n}{-2l+1}{n_2}$ and $\Ifourbub{n}{n_1}{-2l+1}$, respectively, with $l<k$. For example, the following relations of this kind will be useful later, 
\begin{align}\label{eq:Int4bReduce1}
    \Ifourbub{n}{-2}{n_2} &= -2(\delta_1+3)\Ifourbub{n}{-1}{n_2}\,, \nonumber\\
    \Ifourbub{n}{n_1}{-2} &= -2(\delta_2+3)\Ifourbub{n}{n_1}{-1}\,, \nonumber\\
    \Ifourbub{-2}{n_1}{n_2} &= -2(\delta+3)\Ifourbub{-1}{n_1}{n_2}\,.
\nonumber \\
    \Ifourbub{-4}{n_1}{n_2} &= -4(\delta+5)\Ifourbub{-3}{n_1}{n_2} + 8(\delta+3)(\delta+4)(\delta+5)\Ifourbub{-1}{n_1}{n_2}\,.
\end{align}

While in previous examples it was convenient to start with the difference equations in the parameters of the Mellin amplitude, external dimensions $\Delta_i$, Mellin variables and internal dimensions $\delta_i$, for the four-point bubble integral they are deeply intertwined with the SBP relations, and it is useful to start by discussing the latter. 

\subsubsection{The \texorpdfstring{$c_1$}{c1}-SBP relation}

To derive the SBP relation for the variable $c_1$ (and by symmetry also for $c_2$), 
we choose the polynomial 
\begin{align}
    P_{c,c_1,c_2} = \prod_{\sigma=\pm 1}(c+h-c_1+\sigma c_2)(c-h+c_1+\sigma c_2)
\end{align}
to cancel the poles of $\Wfourbub(c_1+2)/\Wfourbub(c_1)$ and consider  \begin{equation}
0 = \cint{c_1}D_{c_1}^{+}\Big[P_{c,c_1,c_2}\Pi_n(c-\delta)\Pi_{n_1}(c_1-\delta_1)\Pi_{n_2}(c_2-\delta_2)\Wfourbub(c,c_1,c_2)\Big] \ .
\end{equation}
The SBP relation can be written formally as\footnote{The analytic expressions for the coefficients $a_{\pmb{n}}$ are given in the ancillary file {\tt SBP\_c1.m}.}
\begin{align}\label{eq:c1_SBP_4b}
    0 = \sum_{\pmb{n}} a_{\pmb{n}}^{} \mathcal{I}^{\text{4-bub}}_{\pmb{n}} \,,
\end{align}
where the sum runs over the 25 triplets $\pmb{n}=(n,n_1,n_2)$
\begin{align}
    \pmb{n}&\in\Big\{(-4 + n, 1 + n_1, n_2)\,, (-3 + n, 1 + n_1, n_2)\,, (-2 + n, -1 + n_1, 
  n_2)\,, (-2 + n, n_1, n_2)\,, \nonumber\\
  &\qquad (-2 + n, 1 + n_1, -2 + n_2) \,, (-2 + n, 
  1 + n_1, -1 + n_2)\,, (-2 + n, 1 + n_1, n_2)\,, \nonumber\\
  &\qquad (-1 + n, -1 + n_1, 
  n_2)\,, (-1 + n, n_1, n_2)\,, (-1 + n, 1 + n_1, -2 + n_2)\,,\nonumber\\
  &\qquad (-1 + n, 
  1 + n_1, -1 + n_2)\,, (-1 + n, 1 + n_1, n_2)\,, (n, -3 + n_1, 
  n_2)\,, (n, -2 + n_1, 
  n_2)\,, \nonumber\\
  &\qquad (n, -1 + n_1, -2 + n_2)\,, (n, -1 + n_1, -1 + n_2)\,, (n, -1 + n_1, 
  n_2)\,, (n, n_1, -2 + n_2)\,,\nonumber\\
  &\qquad (n, n_1, -1 + n_2)\,, (n, n_1, n_2)\,, (n, 
  1 + n_1, -4 + n_2)\,, (n, 1 + n_1, -3 + n_2)\,, \nonumber\\
  &\qquad (n, 1 + n_1, -2 + n_2)\,, (n,
   1 + n_1, -1 + n_2)\,, (n, 1 + n_1, n_2)\Big\}\,.
   \label{eq:16integrals}
\end{align}
It turns out that one choice of indices, $(n,n_1,n_2)=(-1,1,1)$, is particularly useful. In this case, certain coefficients $a_{\pmb{n}}$ vanish. Together with \cref{eq:4bub_zeroes}, we are left with a linear combination of the following 16 integrals:
\begin{align}\label{eq:IntToReduce}
    & \Ifourbub{-5}{2}{1}(s) & &\underline{\Ifourbub{-4}{2}{1}(s)} & &\Ifourbub{-3}{1}{1}(s) & & \uwave{\Ifourbub{-3}{2}{-1}(s)} & &
\Ifourbub{-3}{2}{1}(s) & & \underline{\Ifourbub{-2}{1}{1}(s)}  \nonumber\\
    & \underline{\Ifourbub{-2}{2}{-1}(s)} & & \underline{\Ifourbub{-2}{2}{1}(s)} & &
\underline{\Ifourbub{-1}{-2}{1}(s)} & & \Ifourbub{-1}{-1}{1}(s) & & \Ifourbub{-1}{1}{-1}(s) & &
\Ifourbub{-1}{1}{1}(s) \nonumber\\ 
    &\Ifourbub{-1}{2}{-3}(s) & & \underline{\Ifourbub{-1}{2}{-2}(s)} & & \Ifourbub{-1}{2}{-1}(s) & & \Ifourbub{-1}{2}{1}(s)  \ .
\end{align}
This set of integrals can be further reduced by using \cref{eq:Int4bReduce1} to eliminate all the integrals with a negative even index (those with an underline).

A second choice of indices in \cref{eq:c1_SBP_4b} that is useful for us is 
$(n,n_1,n_2)=(-1,-1,2)$ and $(n,n_1,n_2)=(-1,2,-1)$. The second set of indices may 
be obtained from the former by interchanging $\delta_1$ with $\delta_2$. 
This choice of indices yields a relation for the reduction of $\Ifourbub{-3}{2}{-1}$, with a wavy underline in \cref{eq:16integrals},
\begin{align}
\label{eq:secondc1SBP}
    \Ifourbub{-3}{2}{-1} &= \Ifourbub{-1}{2}{-3} - 2(\delta_1-1)\Ifourbub{-1}{1}{-1} \nonumber\\
    &\quad + (h^2-4+24\delta+3\delta^2+4\delta_1-\delta_1^2-24\delta_2-3\delta_2^2)\Ifourbub{-1}{2}{-1}\,.
\end{align}
Last but not least, a bit of foresight suggests that the (reduction of the) integral $\Ifourbub{-1}{2}{1}(s)$ is also of interests to us, because this integral appears in the difference equation for $\delta_1$,
\begin{align}\label{eq:DE_delta1}
    D_{\delta_1}^{-}\Big[\Ifourbub{-1}{1}{1}(\delta_1,s)\Big] = - \Ifourbub{-1}{2}{1}(\delta_1,s)\,.
\end{align}
Therefore, our next goal is to reduce the other integrals in \cref{eq:IntToReduce} into a linear combination of
\begin{align}\label{eq:4bub_basis}
    \Ifourbub{-1}{2}{1}\,,\qquad\Ifourbub{-1}{1}{1}\,,\qquad\Ifourbub{-1}{1}{-1}\,,\qquad\Ifourbub{-1}{-1}{1}\,.
\end{align}
Because of the negative index in the third and second position, respectively, the last two integrals here can be considered as boundary terms, as we expect that they are related to tadpole integrals. 
The two identities
\begin{align}
    \cint{c_2}\,c_2\,\Wfourbub(c,c_1,c_2,s) &= -\frac{\Gamma(h)^2}{\Gamma(2h)}\Wtadpole(c_1)\Wscalar(c,s)\,,
    \\
 \cint{c_2}\,c_2^3\,\Wfourbub(c,c_1,c_2,s) 
    & = -\frac{\Gamma(h)^2}{\Gamma(2h)}(c^2+c_1^2-h^2)\Wtadpole(c_1)\Wscalar(c,s) \ ,
\nonumber    
\end{align}
which can be proven by direct integration, help us verify this expectation. Indeed, by further integrating them over $c$ and $c_1$ leads to 
\begin{align}
   \label{eq:negC2reduction}
   & \Ifourbub{-1}{n_1}{-1}(s,\delta_1,\delta_2) = -\frac{\Gamma(h)^2}{\Gamma(2h)} \,\Iscalar{-1}(s) \Itadpole{n_1}(\delta_1)\,,
    \\
    \label{eq:negC2reduction2}
    &\Ifourbub{-1}{n_1}{-3}(s,\delta_1,\delta_2) = - \frac{\Gamma(h)^2}{\Gamma(2h)}\Iscalar{-1}(s)\Big[\Itadpole{n_1-2}(\delta_1) + 2(\delta_1-2n_1+3)\Itadpole{n_1-1}(\delta_1) \\
    &\quad + \big(4n_1^2-8n_1-2hs+4(1-n_1)\delta_1+\delta_1^2+3(\delta_2+4)^2+s\Delta_{1234}-\Delta_{12}\Delta_{34}\big)\Itadpole{n_1}(\delta_1)\Big]\, .\nonumber  
\end{align}
We note that, as expected, the right-hand side of these relations feature tadpole integrals. Moreover, $\Itadpole{n_1}$ can be further reduced to $\Itadpole{1}$ with \cref{eq:tadpoleSBP} and $\Iscalar{-1}$ is given in \cref{eq:Im1scalar}. 
The relations we discussed allow us to reduce all integrals in \cref{eq:16integrals} except
\begin{align}
    \Ifourbub{-5}{2}{1}(s)\,,\qquad\Ifourbub{-3}{2}{1}(s)\,,\qquad\Ifourbub{-3}{1}{1}(s)\,,
\end{align}
to the basis~\eqref{eq:4bub_basis}. 

To reduce these remaining integrals we use the identity 
\begin{align}\label{eq:cn_Int}
    &\cint{c} c^{2k+1} \Wfourbub(c,s) =  \cint{c} \,c^{2k-1} \Big[ (s-h)^2\Wfourbub(c,s)-4\Wfourbub(c,s-2)\Big] \,,
\end{align}
which we prove in the \cref{identity_D12_action} based on the action of the conformal Casimir $\mathcal{D}_{12}$ on Mellin-space correlators.
Repeatedly applying this relation until the power of $c$ is reduced to unity and integrating the result with the measure $\int \prod_{i=1}^2 \dd c_i \, \Pi_{n_i}(c_i-\delta_i)$ leads to  
\begingroup
\allowdisplaybreaks
\begin{align}
\label{eq:Innreduction}
    \Ifourbub{-3}{n_1}{n_2}(s) &= \Big[(s-h)^2 + 3\delta^2+24\delta+44\Big]\Ifourbub{-1}{n_1}{n_2}(s) - 4 \,\Ifourbub{-1}{n_1}{n_2}(s-2)\,,\\
    \Ifourbub{-5}{n_1}{n_2}(s) &= \Big[(s-h)^4+10(\delta^2+12\delta+34)(s-h)^2 \nonumber\\*
    &\quad +(5\delta^4+120\delta^3+1020\delta^2+3600\delta+4384)\Big]\Ifourbub{-1}{n_1}{n_2}(s) \nonumber\\*
    &\quad -8(s^2-2hs-2s+h^2+2h+5\delta^2+60\delta+172)\Ifourbub{-1}{n_1}{n_2}(s-2)
    \nonumber\\*
    &\quad +16\Ifourbub{-1}{n_1}{n_2}(s-4) \, ,
\label{eq:Innreduction2}    
\end{align}
\endgroup
which indeed express the desired integrals in terms of integrals in which the auxiliary propagator is collapsed. 
The peculiar feature of these relations is that they trade powers of the auxiliary spectral variable $c$ for shifts in the Mandelstam variable. This is the origin of the intertwining of the difference equations in external variables and the SBP relations anticipated in the beginning of this section.   

Putting together the $c_1$-SBP relation for $(n, n_1, n_2) = (-1, 1, 1)$, \cref{eq:secondc1SBP,eq:negC2reduction,eq:negC2reduction2,eq:Innreduction,eq:Innreduction2}, 
we obtain a relation of the form
\begin{align}
\label{eq:R1}
    0 = \mathcal{R}_1(s) & \equiv \sum_{n=0}^{2}\alpha_{-1,2,1}(n)\Ifourbub{-1}{2}{1}(s-2n) + \sum_{n=0}^{1}\alpha_{-1,1,1}(n)\Ifourbub{-1}{1}{1}(s-2n) \nonumber\\
    &\quad + \alpha_{-1,1,-1}\Ifourbub{-1}{1}{-1}(s) + \alpha_{-1,-1,1}\Ifourbub{-1}{-1}{1}(s) \,,
\end{align}
and the coefficients are given by
\begingroup
\allowdisplaybreaks
\begin{align}
\alpha_{-1,-1,1} &= 2 (\delta_{1}-1) (2 h-1)\,,\nonumber\\
\alpha_{-1,1,-1} &= \frac{(\delta_1-1) (2 h-1) \left[-\Delta_{12} \Delta_{34}+s \Delta_{1234}-\delta_1^2-\delta_2^2+2 \delta_1+2 h^2-2 h (s+1)\right]}{(\delta_1+h-2) (\delta_1+h-1)}\,,\nonumber\\
\alpha_{-1,1,1}(1) &= 16 (\delta_1-1) (h-1)\,, \nonumber\\
\alpha_{-1,1,1}(0) &= 4 (\delta_1-1) \left[-\delta_1^2+\delta_2^2+2 \delta_1+2 h^2 (s-1)\right.\nonumber\\*
&\qquad\qquad\qquad\left.-h \left(-\delta_1^2+\delta_2^2+2 \delta_1+s^2+2
   s-4\right)+s^2-2\right]\,,\nonumber\\
   \alpha_{-1,2,1}(2) &= -16\,,\nonumber\\
   \alpha_{-1,2,1}(1) &= -8 \left(\delta_1^2+\delta_2^2-4 \delta_1+2 h (\delta_1+s-3)-s^2+2 s+2\right)\,,\nonumber\\
   \alpha_{-1,2,1}(0) &= -\left[\delta_1^2-\delta_2^2-4 \delta_1+s^2+2 (\delta_1-2) s+4\right] \nonumber\\*
   &\quad\quad\times\left[\delta_1^2-\delta_2^2-4 \delta_1+4 h^2-4 h (-\delta_1+s+2)+s^2-2 (\delta_1-2) s+4\right]\,.
\end{align}
\endgroup
We have thus obtained a relation between $\Ifourbub{-1}{1}{1}$ and $\Ifourbub{-1}{2}{1}$ with various shifts in their arguments. The next step is therefore is to find another relation of a similar type, so that we can eliminate $\Ifourbub{-1}{2}{1}(s-2n)$ and express $\Ifourbub{-1}{2}{1}(s)$ in terms of $\Ifourbub{-1}{1}{1}$ and tadpole integrals.
To this end we will use the $c$-SBP relation.\footnote{One might expect that a difference equation in $s$ would come from evaluating $D^{-}_s\big[\Ifourbub{-1}{1}{1}\big]$, as in the tree-level scalar exchange case. However, in fact it would just reproduce the relation~\eqref{eq:Innreduction}, which gives no new information. For the currect problem, the independent difference equation in $s$ actually comes from the SBP relation in $c$.}

\subsubsection{The \texorpdfstring{$c$}{c}-SBP relation}

The starting point of the SBP relation for $c$  is $D_{c}^{+}\Big[\widetilde{P}_{c,c_1,c_2}\Pi_{n}\Pi_{n_1}\Pi_{n_2}\Wfourbub\Big]$,where we choose
\begin{align}
    \widetilde{P}_{c,c_1,c_2} &= -(c-1+h)(c-2+h)(c+s-h)(c-\Delta_{12}+h)(c-\Delta_{34}+h) \nonumber\\
    &\quad\times \prod_{\sigma_1,\sigma_2=\pm 1}(c-h+\sigma_1 c_1+\sigma_2 c_2)
\end{align}
to cancel the poles brought by the action of difference operator on $\Wfourbub$. For generic choices of indices, the resulting SBP relation\footnote{The analytic expressions for the coefficients $b_{\pmb{n}}$ are given in the ancillary file {\tt SBP\_c.m}.}
\begin{align}
    0 = \sum_{\pmb{n}}b_{\pmb{n}}^{}\mathcal{I}^{\text{4-bub}}_{\pmb{n}}
\end{align}
involves 103 terms. With the goal of deriving a second relation between $\Ifourbub{-1}{1}{1}$ and $\Ifourbub{-1}{2}{1}$ we start from the triplets of indices $(n,n_1,n_2)=(0,2,1)$ and $(n,n_1,n_2)=(0,1,1)$ for which the $c$-SBP relations involve the desired integrals and integrals that appeared earlier in our discussion. 
After repeatedly using \cref{eq:cn_Int} and the reduction of tadpole integrals, the above two choices lead, respectively, to the following two relations:
\begin{align}
    0 = \mathcal{R}_2(s) &\equiv \sum_{n=0}^{4}\beta_{-1,2,1}(n)\Ifourbub{-1}{2}{1}(s-2n) + \sum_{n=0}^{3} \beta_{-1,1,1}(n) \Ifourbub{-1}{1}{1}(s-2n) \nonumber\\*
    &\quad + \beta_{-1,1,-1}\Ifourbub{-1}{1}{-1}(s) + \beta_{-1,-1,1}\Ifourbub{-1}{-1}{1}(s)\,, 
    \label{eq:R2}
    \\
    \label{eq:DE_4bub_s}
    0 = \mathcal{R}_3(s) &\equiv \sum_{n=0}^{4} \rho_{-1,1,1}(n)\Ifourbub{-1}{1}{1}(s-2n) + \rho_{-1,1,-1}\Ifourbub{-1}{1}{-1}(s) + \rho_{-1,-1,1}\Ifourbub{-1}{-1}{1}(s)\, .
\end{align}
The relation $\mathcal{R}_3$ can be viewed as a difference equation for $s$, analogous in spirit, for example, to \cref{eq:DE_s_scalar}. For this more complicated problem however, the difference equation is of a higher order. The coefficients in $\mathcal{R}_3(s)$ are given in the ancillary file {\tt Relation\_R3.m}.

Using \cref{eq:negC2reduction} it is straightforward to turn $\mathcal{R}_3(s)$ into a difference equation in the Mellin variable $s$ for Mellin amplitudes~\eqref{eq:M4bub_ME}. Moreover, since the Mellin amplitude is invariant under 
\begin{align}\label{eq:stoDelta12}
    \Delta_{12} \rightarrow 2h - s \qquad s\rightarrow 2h - \Delta_{12}\,,
\end{align}
we can use this transformation to obtain an independent difference equation for boundary weight $\Delta_{12}$. To finalize the difference equation in $\delta_1$, \cref{eq:DE_delta1}, we must still express $\Ifourbub{-1}{2}{1}$ in terms of 
$\Ifourbub{-1}{1}{1}$.

To this end, we note that $\mathcal{R}_1(s)$ and $\mathcal{R}_2(s)$ are two linearly independent relations both involving $\Ifourbub{-1}{2}{1}(s-2n)$. Thus we can use them to solve $\Ifourbub{-1}{2}{1}$ in favor of $\Ifourbub{-1}{1}{1}$. 
The strategy is to recursively eliminate $\Ifourbub{-1}{2}{1}$ in one relation by subtracting it from the second relation evaluated on a shifted value of $s$.
The relation $\mathcal{R}_1(s)$ in \cref{eq:R1} contains only $\Ifourbub{-1}{2}{1}(s-4), \dots, \Ifourbub{-1}{2}{1}(s)$, while the relation $\mathcal{R}_2(s)$ in \cref{eq:R2} also contains $\Ifourbub{-1}{2}{1}(s-8)$ and $\Ifourbub{-1}{2}{1}(s-6)$.
We may choose coefficients $r_2$ and $r_4$ such that the combination 
\begin{align}
   0 = \tilde{\mathcal{R}}_2(s) &= \mathcal{R}_2(s)-r_4\mathcal{R}_1(s-4)-r_2\mathcal{R}_1(s-2)-r_0\mathcal{R}_1(s) \nonumber\\
    &= \sum_{n=0}^{1} \gamma_{-1,2,1}(n)\Ifourbub{-1}{2}{1}(s-2n) + \sum_{n=0}^{3} \gamma_{-1,1,1}(n)\Ifourbub{-1}{1}{1}(s-2n) \nonumber\\
    &\quad + \gamma_{-1,1,-1}\Ifourbub{-1}{1}{-1}(s) + \gamma_{-1,-1,1}\Ifourbub{-1}{-1}{1}(s)
\end{align}
is free of both $\Ifourbub{-1}{2}{1}(s-8)$, $\Ifourbub{-1}{2}{1}(s-6)$ and $\Ifourbub{-1}{2}{1}(s-4)$.  Then, we can choose ${\tilde r}_2$ and ${\tilde r}_0$
so that the combination 
\begin{align}
   0= &\mathcal{R}_1(s)-{\tilde r}_2{\tilde {\mathcal{R}}}_2(s-2)
   -{\tilde r}_0{\tilde {\mathcal{R}}}_2(s) 
\end{align}
is free of both $\Ifourbub{-1}{2}{1}(s-4)$ and $\Ifourbub{-1}{2}{1}(s-2)$. Thus, we can solve it for $\Ifourbub{-1}{2}{1}(s)$ in terms of $\Ifourbub{-1}{1}{1}(s)$ with various arguments. The result is
\begin{align}\label{eq:I4bub21_reduce}
    \Ifourbub{-1}{2}{1}(s) &= \sum_{n=0}^{3} \lambda_{-1,1,1}(n)\Ifourbub{-1}{1}{1}(s-2n) + \lambda_{-1,1,-1}\Ifourbub{-1}{1}{-1}(s) + \lambda_{-1,-1,1}\Ifourbub{-1}{-1}{1}(s) \,,
\end{align}
where the coefficients are given in the ancillary file {\tt SBP\_c1\_Final.m}. 
We note that in writing this expression we used $\mathcal{R}_3(s)$ in \cref{eq:DE_4bub_s} to eliminate $\Ifourbub{-1}{1}{1}(s-8)$ in favor of $\Ifourbub{-1}{1}{1}(s-2k)$ with $k=0,1,2,3$ and other known integrals.
We then plug \cref{eq:I4bub21_reduce} into the difference equation~\eqref{eq:DE_delta1}, and we get the final result for the difference equation in $\delta_1$,
\begin{align}\label{eq:DE_delta1_final}
    \Ifourbub{-1}{1}{1}(\delta_1-2,s) &= \Big[1-2\lambda_{-1,1,1}(0)\Big]\Ifourbub{-1}{1}{1}(\delta_1,s) - 2 \sum_{n=1}^{3}\lambda_{-1,1,1}(n)\Ifourbub{-1}{1}{1}(\delta_1,s-2n) \nonumber\\
    &\quad - 2\lambda_{-1,1,-1}\Ifourbub{-1}{1}{-1}(\delta_1,s) - 2\lambda_{-1,-1,1}\Ifourbub{-1}{-1}{1}(\delta_1,s)\, .
\end{align}
To summarize, \cref{eq:DE_delta1_final} provides a difference equation in $\delta_1$ (and also $\delta_2$ after relabeling). We also have $\mathcal{R}_3(s)$ in \cref{eq:DE_4bub_s} as a difference equation in $s$, and it gives another difference equation in $\Delta_{12}$ after the transformation~\eqref{eq:stoDelta12}. This completes the information we can get from difference equations.

Finally, we note that the $c$ and $c_1$ SBP relations discussed above can also be directly applied to study the four-point bubble integral~\eqref{eq:4bubaux} with an additional bulk-bulk propagator. It corresponds to $\Ifourbub{1}{1}{1}$, and the four-point bubble $\Ifourbub{-1}{1}{1}$ studied in detail here will become a boundary term for the difference equation satisfied by $\Ifourbub{1}{1}{1}$. 
For example, the action of the $s$ difference operator on $\Ifourbub{1}{1}{1}$ relates this integral only to itself and to $\Ifourbub{-1}{1}{1}$. Another example is the action of the $\delta$ difference operator on $\Ifourbub{1}{1}{1}$: it yields $\Ifourbub{2}{1}{1}$ which in turn reduces it back to $\Ifourbub{1}{1}{1}$, $\Ifourbub{-1}{1}{1}$ and tadpole integrals upon use of the $c$ SBP with the triplet of indices $\pmb{n} = (1, 1, 1)$.

\subsection{Consistency checks for four-point bubble}

Let us briefly mention checks of the difference equations using known analytic results. If the internal and boundary weights are the same, $\Delta_i=\Delta'=\Delta$, the Mellin amplitude~\eqref{eq:M4bub_ME} can be written as an infinite sum over conformal partial waves~\cite{Fitzpatrick:2011dm}. If we further resctrict to $h=1$ and $\Delta=2$, the Mellin amplitude has the analytic expression~\cite{Aharony:2016dwx}
\begin{align}\label{eq:M4bub2d}
    \mathcal{M}(s) = -\frac{1}{12\pi}\left[\frac{{}_3F_2(1,1,2-\frac{s}{2};\frac{5}{2},3-\frac{s}{2};1)}{s-4}+\frac{3}{10}\frac{{}_3F_2(2,2,3-\frac{s}{2};\frac{7}{2},4-\frac{s}{2};1)}{s-6}\right]\,.
\end{align}
Since all dimensions are fixed, we may therefore verify only the difference equation in the Mellin variable.
For this choice of boundary and operator dimension, \cref{eq:DE_4bub_s} gives the following relation for Mellin amplitudes:
\begin{align}
    0 &= 512\pi\Gamma^2\Big[6-\frac{s}{2}\Big]\mathcal{M}(s-8) - 128\pi(4s^2-43s+140)\Gamma^2\Big[5-\frac{s}{2}\Big]\mathcal{M}(s-6) \nonumber\\
    &\quad + 32\pi(6s^4-105s^3+738s^2-2408s+3040)\Gamma^2\Big[4-\frac{s}{2}\Big]\mathcal{M}(s-4) \nonumber\\
    &\quad - 8\pi(4s^6-81s^5+700s^4-3268s^3+8608s^2-12032s+6912)\Gamma^2\Big[3-\frac{s}{2}\Big]\mathcal{M}(s-2) \nonumber\\
    &\quad + 2\pi (s-4)^3(s-2)^2(s^2-3s+2)s\Gamma^2\Big[2-\frac{s}{2}\Big]\mathcal{M}(s) + 48(s-4)^3\Gamma^2\Big[2-\frac{s}{2}\Big]\, .
\end{align}
The expression~\cite{Aharony:2016dwx} for the four-point bubble $h=1$ and $\Delta=2$ reproduced in \cref{eq:M4bub2d} indeed satisfies this relation. Note that one may opt to cancel the Gamma functions in this relation, and the coefficients will be polynomials in $s$.

We note that if the analytic expression for $\Ifourbub{-1}{1}{1}(\delta_1,s)$ is known, we may use \cref{eq:DE_delta1_final} to derive an analytic expression for $\Ifourbub{-1}{1}{1}(\delta_1-2,s)$. It is also possible to increase $\delta_1$ by two units.
We can use a linear combination of the $s$ difference equation $\mathcal{R}_3(s)$ in \eqref{eq:DE_4bub_s} and \cref{eq:DE_delta1_final} to derive a new relation expressing $\Ifourbub{-1}{1}{1}(\delta_1,s)$ as a linear combination of $\Ifourbub{-1}{1}{1}(\delta_1-2,s-2n)$. After renaming the variable $\delta_1-2\rightarrow\delta_1$, we get 
\begin{align}\label{eq:DE_delta1_2}
    0 = \Lambda\, \Ifourbub{-1}{1}{1}(\delta_1+2,s) &+ \sum_{n=1}^{3}\tilde{\lambda}_{-1,1,1}(n)\Ifourbub{-1}{1}{1}(\delta_1,s-2n) \nonumber\\
    & + \tilde{\lambda}_{-1,1,-1}\Ifourbub{-1}{1}{-1}(\delta_1,s) + \tilde{\lambda}_{-1,-1,1}\Ifourbub{-1}{-1}{1}(\delta_1,s)\,.
\end{align}
When $\Lambda\neq 0$, we can use this relation to solve $\Ifourbub{-1}{1}{1}(\delta_1+2,s)$ from $\Ifourbub{-1}{1}{1}(\delta_1,s)$.

It is tempting to choose $\delta_1=1$ and attempt to derive from \cref{eq:M4bub2d,eq:DE_delta1_2} an analytic expression for the four-point bubble integral
\begin{align}\label{eq:M4bub_new}
    \Ifourbub{-1}{1}{1}(\Delta_{12}=\Delta_{34}=4,\delta_1=3,\delta_2=1,s)=\vcenter{\hbox{\begin{tikzpicture}[every path/.style={very thick}]
        \pgfmathsetmacro{\r}{1.5};
        \draw (0,0) circle (\r);
        \draw (0,0) circle (\r/3);
        \node at (0,\r/3) [above=0pt] {$\Delta=4$};
        \node at (0,-\r/3) [below=0pt] {$\Delta'=2$};
        \filldraw (\r/3,0) circle (1pt) (-\r/3,0) circle (1pt);
        \draw (\r/3,0) -- (30:\r) node [right=0pt] {$\Delta_3=2$};
        \draw (\r/3,0) -- (-30:\r) node [right=0pt] {$\Delta_4=2$};
        \draw (-\r/3,0) -- (150:\r) node [left=0pt] {$\Delta_1=2$};
        \draw (-\r/3,0) -- (-150:\r) node [left=0pt] {$\Delta_2=2$};
    \end{tikzpicture}}}
\end{align}
at $d=2$.
It turns out however that the coefficient $\Lambda$ vanishes, $\Lambda=0$ for this choice of conformal weights, and \cref{eq:DE_delta1_2} merely gives another relation between the Mellin amplitude~\eqref{eq:M4bub2d},
\begin{align}
    0 &= 2\pi \Gamma^2\Big[2-\frac{s}{2}\Big]\mathcal{M}(s) - \frac{8\pi(3s^3-30s^2+104s-108)\Gamma^2\big[3-\frac{s}{2}\big]\mathcal{M}(s-2)}{s(s-1)(s-2)(s-4)^2} \nonumber\\
    & \quad + \frac{32\pi(3s^2-27s+70)\Gamma^2\big[4-\frac{s}{2}\big]\mathcal{M}(s-4)}{s(s-1)(s-2)(s-4)^3} - \frac{128\pi \Gamma^2\big[5-\frac{s}{2}\big]\mathcal{M}(s-6)}{s(s-1)(s-2)(s-4)^3} \nonumber\\
    & \quad + \frac{6\Gamma^2\big[2-\frac{s}{2}\big]}{s(s-1)(s-2)(s-4)} \,,
\end{align}
which is again exactly satisfied by \cref{eq:M4bub2d}.

On the other hand, it is still possible to derive \cref{eq:M4bub_new} from the difference equation~\eqref{eq:DE_delta1_2}. One can use instead $\delta_1=1+\epsilon$ and take the limit $\epsilon\rightarrow 0$ after $\Ifourbub{-1}{1}{1}(\delta_1+2)$ is obtained. It thus requires us to know the analytic behavior of the bubble~\eqref{eq:M4bub2d} in the neighborhood $\delta_1=1+\epsilon$.

\section{Discussion}\label{sec:discussion}

The AdS/CFT correspondence relates tree-level bulk correlation functions to planar CFT boundary correlators; moreover, loop-level bulk correlators describe non-planar corrections to boundary correlation functions. 
In this paper we introduce a systematic procedure to derive difference equations for the basis integrals in tree-level and one-loop correlation functions, together with a summation-by-parts reduction to the basis.
Our approach extends to AdS space powerful methods developed for flat space loop integrals, such as the 
IBP reduction to master integrals~\cite{Chetyrkin:1981qh, Laporta:2000dsw, Maierhofer:2017gsa, Smirnov:2008iw,  Smirnov:2019qkx} 
and differential equations for determining analytic expressions for the master integrals~\cite{Kotikov:1990kg, Bern:1993kr, Remiddi:1997ny, Gehrmann:1999as}.

To illustrate our construction, which applies to general theories and does not rely on special properties such as supersymmetry, we discussed one tree-level and several one-loop examples, the latter of bubble topology with various numbers of external legs. 
In some of them, we used the derived difference equations together with physical conditions to compute previously-unknown integrals, such as the one-loop two-point bubble integral with distinct internal masses. 
In others we demonstrated the validity of the difference equations by verifying that they are obeyed by the known expressions for the integrals, such as the $h=1$, $\Delta=2$ four-point bubble of Ref.~\cite{Aharony:2016dwx}.
The appearance of difference equations rather than the more standard differential equations governing flat space master integrals is a reflection of the discreetness of the spectrum of quadratic operators in AdS space.

Typically, AdS integrals are given as linear combinations of hypergeometric (or more complicated) functions. In contrast, for special values of the external parameters, the difference equations sometimes uniquely fix them in terms of the inhomogeneous term in the equation, which in our examples was a rational function or a rational combination of Gamma function. 
This observation points to interesting identities for (linear combinations of) hypergeometric functions at unit argument.
We illustrated this observation in sections~\ref{sec:scalar} and \ref{sec:bub2}, where we derived and verified several such identities.

An interesting departure from flat-space intuition is the need to impose physical 
conditions to obtain a unique solution to the different equations. Indeed, as discussed in \cref{sec:tadpole,sec:1loop}, difference equations determine their solutions up to periodic functions of their arguments. Further physical conditions must be imposed, to select physical solutions from the unphysical ones.\footnote{This is analogous in spirit with 2d integrable models, for which the Yang-Baxter equation determines the scattering matrix up to an overall phase which is determined (or at least constrained) by the crossing relation, positivity of the residues at physical poles, etc. }
In \cref{sec:bub2} we suggested several suitable physical conditions which constrain correlation functions given by integrals of bubble topology. Given the mathematical richness of higher-loop and higher-point Feynman diagrams, we expect a similar richness for higher-loop and higher-point Witten diagrams. 
It would be important to understand the complete set of such physical conditions constraining them including, for example, the action of conformal Casimir operators.
%
Moreover, while we have not discussed in detail strategies for solving the ensuing difference equations, it would be interesting to understand if they have a suitable definition analogous with the canonical form of differential equations~\cite{Henn:2013pwa} for flat space Feynman integrals.

A physically-motivated approach to solving the $s$-channel difference equations for multi-point bubble integrals, which makes use of its expected pole structure, is to search for solutions of the form
\begin{align}
{\cal M}(s) \sim \sum_{n, m} \frac{R_{n, m}}{s - (\Delta_n+m)} \ ,
\end{align}
as was done at tree level in \cite{Li:2023azu}.
Here $\Delta_n$ are the dimensions of the primaries that appear in the operator-product expansion of the boundary operators while $m$ label their descendants. The $s$ difference equation then becomes a recursion relation for the residues $R_{n, m}$ which may also constrain the possible values $\Delta_n$. The residues and dimensions are further constrained by difference equations in the other parameters. It would be interesting to explore the conditions under which the number of primaries is finite, as e.g. in cases when the boundary theory is a minimal model.

As presented in \cref{sec:genstrat}, writing Mellin amplitudes as integrals over a potential function was an important part of our construction. We also assumed that the potential function was a ratio of Euler Gamma functions. 
At the price of introducing auxiliary integration variables similar to the discussion of the four-point bubble integral in \cref{sec:4bubble}, the potential for the three- and four-point triangle integral has this general form~\cite{Cardona:2017tsw,Yuan:2018qva}. It would be extremely interesting to compute these integrals, either directly or via difference equations method described here. 
More generally, it would be useful to extend our analysis and relax the assumption that the potential function is a ratio of Gamma functions. Since some knowledge about its structure is necessary, one may e.g. assume that the potential is some more general Euler integral, which still obeys certain difference relations~\cite{Matsubara-Heo:2023ylc}. 
One could also convert the Mellin amplitudes back to the position space such that the difference equations will become differential equations. While these differential equations might be rather complicated, there exist various tools to tackle them, and the periodic function ambiguity in Mellin amplitudes can also be avoided.

Many of the existing loop-level computations of AdS boundary correlators in AdS space have been carried out in specific theories, perhaps with certain amount of supersymmetry. 
With the novel abilities to evaluate --- or at least constrain --- AdS integrals provided by the SBP relations and difference equations, extended to integrals with nontrivial numerators, one may explore special properties of correlators that emerge when particular configurations of particles and couplings contribute to the same correlation function.

While we formulated the SBP relations and the difference equations for AdS boundary correlators, its main ingredients --- the existence of split representation for bulk-bulk propagators, potentials that are ratios of Gamma functions up to auxiliary integrals --- exist also in de Sitter space. We expect that, with appropriate modifications, our construction also holds 
in de~Sitter space and for cosmological correlators.

\begin{acknowledgments}
We thank Ofer Aharony, Qu Cao, Lorenz Eberhardt, Song He, Xuhang Jiang, Sebastian Mizera, Allic Sivaramakrishnan, Yichao Tang and Cristian Vergu for valuable discussions.
A.H. is supported by the Simons Foundation.
R.R. and F.T. are supported by the U.S. Department of Energy (DOE) under award number DE-SC00019066.
\end{acknowledgments}


\newpage
\appendix

\section{Conventions}\label{conv}

In this appendix we collect our conventions and details of the embedding space formalism and bulk and boundary integration that are useful throughout the paper.

The bulk-bulk progatator in $\AdS_{d+1}$ satisfies the equation
\begin{equation}\label{e1q}
[\nabla_{X}^{2}-\Delta(\Delta-d)]\,G_{\Delta}(X,Y)=-\delta^{d+1}(X,Y)\,.
\end{equation}
If we take one of the bulk point to the boundary, we get the bulk-boundary propagator
\begin{equation}\label{bBprop}
E_{\Delta}(X,P)
=\frac{C_{\Delta}}{(-2 X\cdot P)^{\Delta}}\,, \qquad C_{\Delta}=\frac{\Gamma(\Delta)}{2\pi^{d/2}\Gamma(\Delta-d/2+1)}\,,
\end{equation}
where $\Delta$ is the conformal weight of the boundary operator located at $P$. 

It is convenient to put the bulk-bulk propagator into the spectral representation
\begin{align}\label{scalarbulktobulk}
G_{\Delta}(X_{1},X_{2})&=\cint{c}\frac{1}{(\Delta-h)^{2}-c^{2}}\,\Omega_{c}(X_{1},X_{2})\,, \nonumber\\
\Omega_{c}(X_{1},X_{2}) &=-2c^{2}\int_{\bAdS} \dd Q \, E_{d/2+c}(X_{1},Q)E_{d/2-c}(X_{2},Q)
\end{align}
where the spectral function $\Omega_c$ satisfies
\begin{align}\label{eq:spec_del}
    \delta^{d+1}(X_{1},X_{2})&=\cint{c}\,\Omega_{c}(X_{1},X_{2}) \,.
\end{align}
Note that by a direct computation, we have
\begin{align}
    \Omega_c(X,X) = \frac{\Gamma(h)}{2\pi^h\Gamma(2h)}\frac{\Gamma(h+c)\Gamma(h-c)}{\Gamma(c)\Gamma(-c)} \,.
\end{align}
This immediately leads to the spectral representation for tadpoles, see \cref{spectraltp}.

The Mellin representation for the boundary correlator $\mathcal{A}(P_i)$,
\begin{align}
    \mathcal{A}(P_i) = \langle\mathcal{O}_{\Delta_1}(P_1)\mathcal{O}_{\Delta_2}(P_2)\ldots\mathcal{O}_{\Delta_n}(P_n)\rangle\,,
\end{align}
is defined as
\begin{equation}\label{eq:MellinAmp}
\mathcal{A}(P_i)= \frac{\pi^{h}}{2}\Gamma\left[\frac{\sum_{i=1}^{n}\Delta_{i}-d}{2}\right]\prod_{i=1}^{n}\frac{C_{\Delta_i}}{\Gamma(\Delta_i)}\int [\dd\delta_{ij}] \, \mathcal{M} (\delta_{ij})\prod_{i<j}^{n}\frac{\Gamma(\delta_{ij})}{P_{ij}^{\delta_{ij}}}\,,
\end{equation}
where $P_{ij}\equiv -2P_i\cdot P_j$, and we call $\mathcal{M}(\delta_{ij})$ the Mellin amplitude.
The Mellin variables $\delta_{ij}$ obey the constraint
\begin{equation}
\delta_{ij}=\delta_{ji}\,, \qquad \sum_{j=1,\,j\neq i}^{n}\delta_{ij}=\Delta_i\,,
\end{equation}
such that there are $n(n-3)/2$ independent variables. It is convenient to define the Mandelstam variables $s_{ij}$ through
\begin{align}
    \delta_{ij} = \frac{\Delta_i+\Delta_j-s_{ij}}{2}\,,
\end{align}
and $s_{ij}$ satisfy the same relations as the flat space Mandelstam variables $s_{ij} = -(k_i+k_j)^2$ with $\sum_{i=1}^{n}k_i=0$ and $k_i^2 = -\Delta_i$. 
We denote the $n(n-3)/2$ independent Mandelstam variables as $\{\mathsf{s}_{k}\}$, which forms a basis for the Mellin variables $\delta_{ij}$. We follow Ref.~\cite{Penedones:2010ue} and define the integration measure as
\begin{align}
    \int [\dd\delta_{ij}] = \int\limits_{-i\infty}^{i\infty}\prod_{k=1}^{n(n-3)/2}\frac{\dd \mathsf{s}_{k}}{2(2\pi i)} \,.
\end{align}
The simplest example of Mellin amplitudes corresponds to the scalar contact diagram, 
\begin{align}\label{contact}
\mathcal{A}_{\text{contact}}\equiv D_{\Delta_1,\Delta_2,\ldots,\Delta_n}=\int_{\AdS} \dd X\,\prod^n_{i=1} E_{\Delta_{i}}(P_{i},X) \,.
\end{align}
Because of the identity
\begin{equation}\label{mellinamp}
\int_{\AdS} \dd X\,\prod^n_{i=1} E_{\Delta_{i}}(P_{i},X)=\frac{\pi^{h}}{2}\,\Gamma\left[\frac{\sum_{i=1}^{n} \Delta_i-d}{2}\right]\,\prod^n_{i=1}\frac{\mathcal{C}_{\Delta_{i}}}{\Gamma(\Delta_{i})} \int [d\mu] \prod_{i<j}^{n}\frac{\Gamma(\delta_{ij})}{P_{ij}^{\delta_{ij}}}\,,
\end{equation}
the Mellin amplitude for the contact diagram is simply
\begin{align}
\mathcal{M}_{\text{contact}}=1 \,.
\end{align}


\section{Two-point bubble SBP relation coefficients}\label{app:sbpcoef}

In this appendix we spell out the coefficients of the SBP relation in \cref{eq:2bubbleSBP}:
\begingroup
\allowdisplaybreaks
\begin{align}\label{eq:atwobubble}
 a_{n_1-3,n_2} &= 4 h-n_1 \\
 a_{n_1-2,n_2} &= -4 \Big[7 h n_1-3 (\delta_{1}+4) h+n_1 (\delta_{1}-2 n_1+3)\Big] \\
 a_{n_1-1,n_2-2} &= 2 n_1-4 h \\
 a_{n_1-1,n_2-1} &= 4 (2 h-n_1) (-\delta_{2}+2 n_2-3) \\
 a_{n_1-1,n_2} &= -2 \Big[2 h^2 (n_1-2 \pmb{\Delta} )+2 h \Big(-3 \delta_{1}^2+\delta_{2}^2-18 \delta_{1}+4 \delta_{2}+\pmb{\Delta} ^2-18
   n_1^2 \nonumber\\*
 &\quad +n_1 (15 \delta_{1}+\pmb{\Delta} +42)+4 n_2^2-4 (\delta_{2}+2) n_2-24\Big) +n_1 \Big(3 \delta_{1}^2-\delta_{2}^2+12 \delta_{1} \nonumber\\*
 &\quad -4 \delta_{2}-\pmb{\Delta} ^2+12 n_1^2-12 (\delta_{1}+2) n_1-4 n_2^2+4 (\delta_{2}+2) n_2+10\Big)\Big] \\
 a_{n_1,n_2-2} &= 4 \Big[h (-\delta_{1}+3 n_1-2)+n_1 (\delta_{1}-2 n_1+1)\Big] \\
 a_{n_1,n_2-1} &= -8 (-\delta_{2}+2 n_2-3) \Big[h (-\delta_{1}+3 n_1-2)+n_1 (\delta_{1}-2 n_1+1)\Big] \\
 a_{n_1,n_2} &= 4 \Big[2 n_1^2 \Big(3 \delta_{1}^2-\delta_{2}^2+6 \delta_{1}-4 \delta_{2}-\pmb{\Delta} ^2+2 h^2+2 h (6 \delta_{1}+\pmb{\Delta} +9)-4
   n_2^2 \nonumber\\*
   &\quad +4 (\delta_{2}+2) n_2\Big)+n_1 \Big(-2 h^2 (\delta_{1}+3 \pmb{\Delta} +1) +h \big(-2 (\delta_{1}+1) \pmb{\Delta} -9 \delta_{1}^2+3 \delta_{2}^2 \nonumber\\*
   &\quad -30 \delta_{1}+12 \delta_{2}+3 \pmb{\Delta} ^2+12 n_2^2-12 (\delta_{2}+2) n_2-16\big)+(\delta_{1}+1) \big(-\delta_{1}^2+\delta_{2}^2 \nonumber\\*
   &\quad -2 \delta_{1}+4 \delta_{2}+\pmb{\Delta} ^2+4
   n_2^2-4 (\delta_{2}+2) n_2+2\big)\Big) -4 n_1^3 (3 \delta_{1}+5 h+3) \\*
   &\quad +(\delta_{1}+2) h \left(\delta_{1}^2-\delta_{2}^2+4 \delta_{1}-4 \delta_{2}-\pmb{\Delta} ^2+2 \pmb{\Delta}  h-4 n_2^2+4 (\delta_{2}+2) n_2\right)+8 n_1^4\Big] \nonumber\\
 a_{n_1+1,n_2-4} &= -n_1 \\
 a_{n_1+1,n_2-3} &= 4 n_1 (-\delta_{2}+2 n_2-5) \\
 a_{n_1+1,n_2-2} &= 2 n_1 \Big[\delta_{1}^2-3 \delta_{2}^2-24 \delta_{2}+\pmb{\Delta} ^2+2 h^2-2 h (-\delta_{1}+\pmb{\Delta} +2 n_1)+4 n_1^2-4
   \delta_{1} n_1 \nonumber\\*
   &\quad -12 n_2^2+12 \delta_{2} n_2+48 n_2-50\Big] \\
 a_{n_1+1,n_2-1} &= -4 n_1 (-\delta_{2}+2 n_2-3) \Big[\delta_{1}^2-\delta_{2}^2-6 \delta_{2}+\pmb{\Delta} ^2+2 h^2-2 h (-\delta_{1}+\pmb{\Delta} +2
   n_1) \nonumber\\*
   &\quad +4 n_1^2-4 \delta_{1} n_1-4 n_2^2+4 \delta_{2} n_2+12 n_2-10\Big] \\
 a_{n_1+1,n_2} &= -n_1 \Big[2 \delta_{1} \pmb{\Delta} +\delta_{1}^2-\delta_{2}^2-4 \delta_{2}+\pmb{\Delta} ^2+4 n_1^2-4 n_1 (\delta_{1}+\pmb{\Delta} )-4
   n_2^2 \nonumber\\*
   &\quad +4 (\delta_{2}+2) n_2-4\Big] \Big[-2 \delta_{1} \pmb{\Delta} +\delta_{1}^2-\delta_{2}^2-4 \delta_{2}+\pmb{\Delta} ^2+4 h^2 \\*
   &\quad -4 h (-\delta_{1}+\pmb{\Delta} +2 n_1)+4
   n_1^2+4 n_1 (\pmb{\Delta} -\delta_{1})-4 n_2^2+4 \delta_{2} n_2+8 n_2-4\Big] \nonumber
\end{align}
\endgroup

\section{Derivation of the relation~\texorpdfstring{\eqref{eq:cn_Int}}{(5.57)}}\label{identity_D12_action}

To prove \cref{eq:cn_Int} it is convenient to start 
by rewriting the four-point bubble integral \cref{eq:4bub} as
\begin{align}\label{eq:A4bub_alt}
    \mathcal{A}_{\text{4-bub}} = \int_{\AdS}\dd X_1 \dd X_2 \dd X_3 \, & E_{\Delta_1}(P_1,X_1)E_{\Delta_2}(P_2,X_1)\delta^{(d+1)}(X_1,X_3)G_{\Delta}(X_3,X_2) \nonumber\\
    &\times G_{\Delta'}(X_3,X_2)E_{\Delta_3}(P_3,X_2)E_{\Delta_4}(P_4,X_2)\, ,
\end{align}
and use the spectral representation $\Omega_c(X_1,X_3)$ of the delta-function as given in \cref{eq:spec_del}. We may then follow \cref{sec:4bubble} and derive the explicit expression for $\Wfourbub$, but here we keep the bulk integrals explicit. We define the spectral Mellin amplitude $\mathscr{M}$ as
\begin{align}
    \mathcal{M}_{\text{4-bub}}(s) = \cint{c}\,\mathscr{M}(c,s)\,,
\end{align}
such that $\mathscr{M} \propto c \, \Wfourbub$.
The integral on the left hand side of \cref{eq:cn_Int} is proportional to the integral of $c^{2k}\mathscr{M}$,
\begin{align}\label{eq:MtoW_alt}
    \frac{\cint{c}\, c^{2k+1}\Wfourbub(c,s) }{\pi^h \Gamma(h)\Gamma\left(\frac{\Delta_{1234}}{2}-h\right)\Gamma\left(\frac{\Delta_{12}-s}{2}\right)\Gamma\left(\frac{\Delta_{34}-s}{2}\right)} = \cint{c} c^{2k} \mathscr{M}(c,s)\,.
\end{align}
It is also useful to consider this family of four-point bubble integrals,
\begin{align}
    \mathcal{A}_{\text{4-bub}}^{2k} = \cint{c} \int[\dd\delta_{ij}] c^{2k}\mathscr{M}(c,\delta_{ij})\prod_{i<j}^{4}\frac{\Gamma(\delta_{ij})}{P_{ij}^{\delta_{ij}}}\, ;
\end{align}
the original integral $\mathcal{A}_{\text{4-bub}}$ corresponds to $k=0$. To proceed, we will convert powers of $c^2$ into differential operators acting on the spectral function $\Omega_c$, and then use conformal Ward identities so that these operators eventually act on boundary data.

The detailed calculation uses techniques that resemble those in Ref.~\cite{Fitzpatrick:2011ia}. Let us recall the definition of conformal generators acting on bulk and boundary points,
\begin{align}
D_{X}^{AB}&=\frac{1}{\sqrt{2}}\left ( X^{A}\frac{\partial}{\partial X_{B}}-X^{B}\frac{\partial}{\partial X_{A}} \right )\,, \nonumber\\
D_{i}^{AB}&=\frac{1}{\sqrt{2}}\left ( P_i^{A}\frac{\partial}{\partial P_{i,B}}-P_i^{B}\frac{\partial}{\partial P_{i,A}} \right )\,,
\end{align}
and of their contractions,
\begin{align}
    D_{X}^2 = D_{X}\cdot D_{X}=-\nabla^2_X\,,\qquad \mathcal{D}_{ij} = (D_i + D_j)^2 = D_i^2 + 2D_i\cdot D_j + D_j^2 \,.
\end{align}
The spectral function satisfies the identity
\begin{align}
    c^2\,\Omega_c(X_1,X_3) = -(D_{X_1}^{2}-h^2)\Omega_c(X_1,X_3)\,.
\end{align}
Integration by parts allows us to move the derivatives onto the bulk-boundary propagators,
\begin{align}
    & \int\dd X_1\,E_{\Delta_1}(P_1,X_1)E_{\Delta_2}(P_2,X_1)\Big[D_{X_1}^2 \Omega_c(X_1,X_3)\Big] \nonumber\\
    & = \int\dd X_1\,D_{X_1}^2\Big[E_{\Delta_1}(P_1,X_1)E_{\Delta_2}(P_2,X_1)\Big] \Omega_c(X_1,X_3) \,.
\end{align}
The conformal Ward identity implies that 
\begin{align}
     D_X^2\Big[E_{\Delta_1}(P_1,X)E_{\Delta_2}(P_2,X)\Big] = \mathcal{D}_{12}\Big[E_{\Delta_1}(P_1,X)E_{\Delta_2}(P_2,X)\Big]\,,
\end{align}
such that the derivatives are now acting on boundary points.
Therefore, we have the relation
\begin{align}\label{eq:D12_action}
    & \cint{c}\int[\dd\delta_{ij}] c^{2k}\mathscr{M}(c,s)\prod_{i<j}^4\frac{\Gamma(\delta_{ij})}{P_{ij}^{\delta_{ij}}} \nonumber\\
    & = -(\mathcal{D}_{12}-h^2) \cint{c}\int [\dd\delta_{ij}] c^{2k-2}\mathscr{M}(c,s) \prod_{i<j}^{4}\frac{\Gamma(\delta_{ij})}{P_{ij}^{\delta_{ij}}}\,.
\end{align}
Note that $\mathscr{M}(c,s)$ only depends on the Mandelstam variable $s$, which is related to the Mellin variable $\delta_{12}$ through \begin{equation}
\delta_{12}=\frac{\Delta_{12}-s}{2} \, .
\end{equation}
At four points, there are two independent Mellin variables, and we choose them to be $\delta_{12}$ and $\delta_{23}$. Acting $\mathcal{D}_{12}$ on $1/P_{ij}^{\delta_{ij}}$ will shift the exponents. We can counter-act this effect by shifting the Mellin variables such that the Mellin integral remains in the same form. This procedure will shift the argument in $\mathscr{M}(c,s)$. Working out the algebra on the right hand side, we can rewrite \cref{eq:D12_action} as
\begin{align}
    & \int[\dd\delta_{ij}] c^{2k}\mathscr{M}(c,s)\prod_{i<j}^{4}\frac{\Gamma(\delta_{ij})}{P_{ij}^{\delta_{ij}}} \nonumber\\
    & = \int[\dd\delta_{ij}] c^{2k-2}\Big[(s-h)^2\mathscr{M}(c,s) - (s-\Delta_{12})(s-\Delta_{34})\mathscr{M}(c,s-2)\Big]\prod_{i<j}^{4}\frac{\Gamma(\delta_{ij})}{P_{ij}^{\delta_{ij}}}\,.
\end{align}
Note that $\mathscr{M}(c,s)$ and $\Wfourbub(c,s)$ differ by an overall factor, see \cref{eq:MtoW_alt}. If we change to $\Wfourbub(c,s)$, then the overall factor will absorb the extra factor in the second term of the above equation while $s$ is shifted. This gives us the desired relation
\begin{align}
    \cint{c}\,c^{2k+1} \Wfourbub(c,s) = \cint{c}\,c^{2k-1} \Big[ (s-h)^2 \Wfourbub(c,s) -4 \Wfourbub(c,s-2) \Big]\,,
\end{align}
and we have derived \cref{eq:cn_Int}.

\bibliographystyle{JHEP}
\bibliography{Draft.bib}

\end{document}